\documentclass[twocolumn,twocolappendix]{aastex631}
\usepackage{listings}
\usepackage{multirow}
\usepackage{breqn}
\usepackage{amsmath}
\usepackage[utf8]{inputenc}
\usepackage{amssymb}
\begin{document}

\title{Forecasting constraints from surface brightness fluctuations in galaxy clusters}
\author[0000-0001-5725-0359]{Charles E. Romero}
\altaffiliation{E-mail: \href{mailto:charles.romero@cfa.harvard.edu}{charles.romero@cfa.harvard.edu} }
\affiliation{Center for Astrophysics $\vert$ Harvard \& Smithsonian, 60 Garden Street, Cambridge, MA 02138, USA}

\shorttitle{Forecasting Fluctuations}
\shortauthors{Romero}

\begin{abstract}

    %Sunyaev-Zel'dovich (SZ) observations have been
    Studies of surface brightness (SB) fluctuations in the intracluster medium (ICM) present an indirect estimate of turbulent pressure support and associated Mach numbers. While high resolution X-ray spectroscopy offer means to directly constrain line of sight gas motions, including those due to turbulence, such observations are relatively expensive and will be limited to nearby, bright clusters. In this respect, SB fluctuations are the most economical means to constrain turbulent motions at large cluster radii across a range of redshifts and masses. 
    
    To forecast what current and future X-ray and SZ facilities may achieve in SB fluctuation studies, I review and synthesize matters of accuracy and precision with respect to calculating power spectra of SB fluctuations, from which turbulent properties are derived. Balance concerns of power spectrum accuracy and precision across a range of spatial scales, I propose the use of three annuli with: [0,0.4] $R_{500}$, [0.4,1] $R_{500}$, and [1,1.5] $R_{500}$. Adopting these three regions, I calculate the uncertainty in the hydrostatic mass bias, $\sigma_{b_{\mathcal{M}}}$, can be achieved for various instruments in several scenarios. I find that \textit{Lynx} and AtLAST are competitive in their constraints at $R_{500}$, while AtLAST should perform better when constraining $\sigma_{b_{\mathcal{M}}}$ at $R_{200}$. 
    
\end{abstract}

%\keywords{Galaxy clusters are awesome (001)}
\section{Introduction}
\label{sec:intro}

    Galaxy clusters are filled with hot, ionized gas known as the intracluster medium (ICM). It can be probed by X-ray (via brehmsstrahlung and line emission) and via the Sunyaev-Zel'dovich \citep[][; SZ]{sunyaev1972} effect. These two probes are complementary as X-ray emission in strongly related to gas density and the SZ (specifically the thermal SZ, hereafter tSZ) effect is proportional to the (thermal) electron pressure. From density and pressure distributions we can infer gas temperature and entropy distributions. Despite this wealth of information, key pieces are missing. 

    One key piece missing is velocity information \citep[see][for a review]{simionescu2019}. Gas velocities may be inferred for some substructures \citep[e.g.][]{markevitch2007,sayers2013b,adam2017a,vanweeren2019}, but the gas velocity away from these substructures remains unconstrained. In particular, turbulent motions are expected to fill the ICM volume \citep[][]{simionescu2019}, and turbulence will play an important role in heating, cooling, and gas and energy transport \citep[][]{rebusco2007,chen2019,gaspari2020,wang2023}
    %One open question is how the ICM is heated, from the central regions to outskirts \citep[e.g]{kravtsov2012,chen2019,gaspari2020}. Away from the central regions, we should expect the AGN is subdominant, but we may still ask whether shock (predominantly adiabatic) heating or turbulent (i.e. dissipative) heating is dominant at moderate to large radii. This has not been answered observationally, though simulations suggest turbulent heating should be dominant at $r < R_{500}$\footnote{$R_{500}$ is the radius within which the mean density is 500 times the critical density of the universe.}, while shock heating should dominate at $r > R_{500}$ \citep{shi2020MNRAS}.
    
    In addition to the aforementioned motivations of studying turbulence, a major driver of studying turbulence is assessing its contribution to the non-thermal pressure in galaxy clusters.
    That is, a common approach to estimating cluster masses has been to assume hydrostatic equilibrium with solely thermal pressure support. It is unclear how much non-thermal pressure support derives from turbulent pressure specifically \citep[e.g.][]{battaglia2012a,lau2013,nelson2014a,biffi16,angelinelli2020}. Observationally, many approaches to estimate the hydrostatic mass bias have been taken\citep[e.g.][]{maughan2016,andreon2017,ettori2020,logan2022,wicker2023}, though most credence appears to be afforded to avenues which contrast a hydrostatic mass estimate to a total mass estimate from either weak lensing\citep[e.g.][]{applegate2014,hoekstra2015,sereno2015a,mantz2016,okabe2016,sereno2017,bulbul2024} or cosmic microwave background (CMB) lensing \citep[e.g.][]{hurier2018,madhavacheril2020,ansarinejad2024}. 

    X-ray spectroscopy \citep[e.g.][]{sanders2010,pinto2015,ogorzalek2017,sanders2020}, and especially high resolution X-ray spectroscopy \citep{Hitomi2016,Hitomi2018} will directly constrain gas motions and thus place constraints on turbulent velocities. While their direct access to gas motions is critical, these observations, for \textit{XRISM} \citep[][]{kitayama2014,zuhone2016,ota2018} or \textit{NewAthena} \citep[e.g.][]{roncarelli2018,ZhangC2024}, will still be expensive (potentially prohibitively so), especially if probing out to $R_{500}$ or beyond.

    Yet another avenue to accessing gas motions (indirectly) comes from studies of surface brightness (SB) \citep[e.g.][]{churazov2012} or thermodynamic fluctuations \citep{schuecker2004,hofmann2016,ueda2018}. Initially these studies focused on the central (brightest) regions of galaxy clusters \citep[e.g.][]{Zhuravleva2015,arevalo2016,romero2023}, but deeper observations have allowed studies out to larger radii \citep{eckert2017,zhuravleva2019,sanders2020,heinrich2024}, now with a few studies produce constraints even out to $R_{500}$ \citep[][]{khatri2016,dupourque2023,dupourque2024,romero2024}. While studies of SB fluctuations are still expensive, they offer more affordable access to constraints on turbulent motions, especially in lower density regions of clusters, than high resolution X-ray spectroscopy.

    The utility of surface brightness fluctuations in the intracluster medium (ICM) arises due to the optically thin nature of the ICM via either the Sunyaev-Zel'dovich (SZ) effect or bremsstrahlung radiation in the X-rays. The surface brightness in SZ (specifically the thermal SZ, hereafter tSZ) is characterized by the Compton y parameter: $y \propto \int P_e dz$, where $P_e$ is the thermal electron pressure and $dz$ indicates integration along the line of sight. The analog for X-ray is $I \propto n_e^2 dz$, where $n_e$ is the electron density and the (omitted) temperature dependence is often quite negligible, especially for hot ($k T_e > 3$ keV) ICM gas when observing in energy bands between roughly 1 and 2 keV.

%    One does not strictly need to follow this approach. In particular, one can compute the power spectrum on the absolute (not relative) fluctuations, as was done in \citet{schuecker2004}. Another alternative is to compute the structure function \citep[e.g][]{roncarelli2018,clerc2019,cucchetti2019}. \textcolor{red}{Cite also Zuhone+ 2016, Seta+ 2023?} This paper is not intended to compare these methods, but rather, to explore modes of calculating the 2D power spectra, $P_{\text{2D}}$ of the relative fluctuations.
    The foundation of Fourier analyses of fluctuations in the ICM is the relation between the power spectra of 2-dimensional surface brightness fluctuations, $P_{\text{2D}}$ and the power spectra of the corresponding 3-dimensional thermodynamic quantities, $P_{\text{3D}}$ \citep{peebles1993, churazov2012}:
    \begin{equation}
        P_{\text{2D}}(k_{\theta}) = \int P_{\text{3D}}(\mathbf{k}) |\Tilde{W}|^2 dk_z,
        \label{eqn:formalism_full}
    \end{equation}
    where $k_{\theta}$ is a circular wavenumber (in 2D), $\mathbf{k}$ is a spherical wavenumber in 3D, $k_z$ is the wavenumber with respect to the scales along the line of sight and $\tilde{W}$ is the Fourier transform of the window function, $W$.
    A (surface brightness) fluctuation map is taken as the residual map divided the model of ICM surface brightness to get either a $\delta y/y$ or $\delta S/S$ map. For clarity, I write $\delta y / \bar{y}$ and $\delta S / \bar{S}$ so as to denote that $\bar{y}$ and $\bar{S}$ are models (e.g. fitted radial profiles) of the surface brightness.

    In practice, Equation~\ref{eqn:formalism_full} is often approximated as:
    \begin{equation}
        P_{\text{2D}}(k_{\theta}) \approx P_{\text{3D}}(\mathbf{k}) \int |\Tilde{W}|^2 dk_z,
        \label{eqn:formalism_approx}
    \end{equation}    
    and the integral is assigned its own variable:
    \begin{equation}
        N(\theta) = \int |\Tilde{W}|^2 dk_z.
        \label{eqn:NofTheta}
    \end{equation}

    \subsection{Overview}
    
        \begin{figure}
            \centering
            \includegraphics[width=0.47\textwidth]{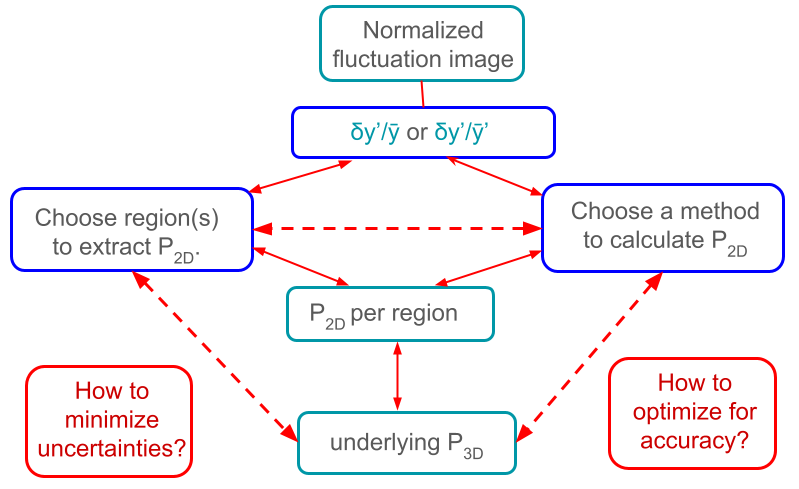}
            \caption{A flowchart which shows the interconnected nature of choices specific to power spectra calculation. Teal boxes indicate data products; blue boxes indicate analysis choices, and red boxes are questions to be asked for each choice to be made. The prime ($^{\prime}$) notation denotes that the respective map is convolved by the instrument beam.}
            \label{fig:flowchart}
        \end{figure}

        With respect to SB fluctuation analyses, Figure~\ref{fig:flowchart} highlights the interconnectedness of data products and analysis choices. 
        In Section~\ref{sec:accuracy} I consider various biases and address their relative importance.  In Section~\ref{sec:uncertainties}, I focus on uncertainties in measured power spectra of SB fluctuations, especially for a circular region and show where a minimum in such uncertainties can be obtained. Section~\ref{sec:synthesis} synthesizes the prior two sections and sets the stage for investigating what constraints on the turbulent hydrostatic mass bias may be achieved with current and future X-ray and SZ instruments (Section~\ref{sec:approach}). Idealized forecasting results are presented in Section~\ref{sec:results} and in Section~\ref{sec:discussion} I discuss limitations and caveats of these forecasts and I consider how biases will reduce the idealized constraints. Conclusions are presented in Section~\ref{sec:conclusions}.

            Throughout this paper, I adopt a 3D power spectrum with the following parameterization:
        \begin{equation}
            P_{\text{3D,model}}(k) = P_0 e^{-(k_{\text{c}}/k)^{\eta_{\rm c}}} k^{- \alpha} e^{-(k/k_{\text{dis}})^{\eta_{\rm d}}},
            \label{eqn:ps_model}
        \end{equation}
        where $P_0$ is the normalization, $k_{\text{c}}$ is a cutoff wavenumber, $\alpha$ is the spectral index, $k_{\text{dis}}$ is a dissipation wavenumber, $\eta_{\rm c}$ and $\eta_{\rm d}$ regulate how quickly the cutoff and dissipation occur. In \citet{Gaspari2013_PS}, approximate values for $\eta_{\rm c}$ and $\eta_{\rm d}$ are found to be 4 and 3/2, respectively. Except for an exploration in biases, I largely neglect the dissipation term in this study.
        Note that if one defines the injection scale as where the amplitude spectrum, $A_{\text{3D}} = \sqrt{4 \pi k^3 P_{\text{3D}}}$ peaks, then this occurs where $d \ln P_{\text{3D}} / d \ln k = -3$. In the case of the model spectrum (and $k_{\text{dis}} \gg k_{\text{c}}$) the injection wavenumber is $k_{\text{inj}} = \eta_{\rm c} k_c / (\alpha - 3)^{1/\eta_{\rm c}}$. For a Kolmogorov spectrum ($\alpha = 11/3$) this would mean $k_{\text{inj}} = (1.5/\eta_{\rm c})^{(1/\eta_{\rm c})} k_c$.

    I adopt a concordance cosmology: $H_0 = 70$~km~s$^{-1}$~Mpc$^{-1}$, $\Omega_M = 0.3$, $\Omega_{\Lambda} = 0.7$. Unless otherwise stated all uncertainties are reported as the standard deviation (for distributions taken to be symmetric) or the distance from the median to the 16th and 84th percentiles (when allowing for asymmetric distributions).

\section{Addressing the accuracy of power spectra of fluctuations}
\label{sec:accuracy}

    In this section I address the matter of accuracy by investigating several biases. Of principal concern are biases due to the method of power spectrum estimation and physically induced biases for a perfectly spherical cluster. Other biases, such as those due to gas clumping, cluster geometry, and substructure \citep[e.g.][]{dupourque2023,heinrich2024} are better investigated suited to investigation with hydrodynamical or cosmological simulations \citep[e.g.][]{zhuravleva2023}, which are beyond the scope of this work.
    
    \subsection{Choice of Estimator}
    \label{sec:PS_estimator}

        In the introduction I already indicated a focus on the power spectra of normalized fluctuation maps. There are direct approaches to estimating the Fourier transform or power spectrum of a region within an image \citep[e.g.][]{koch2019}, several of which are designed to handle arbitrary masking, such a delta-variance methods \citep{ossenkopf2008,arevalo2012}. When a map is mostly empty one can employ a structure function \citep[e.g][]{zuhone2016,seta2023}, or even if a map is filled but the distances between cells is irregular as  with Voronoi tessellation \citep{roncarelli2018,ZhangC2024}. Although much of the work of studies of fluctuations is done for power spectra of a flat-sky image, if one is working with a spherical sky (e.g. a HEALpix map: \citet{gorski2005}), the spherical harmonic transforms (or Pseudo-$C_\ell$s) can be calculated via packages like PolSpice \citep{chon2004}, NaMaster \citep{alonso2019}, or anafast \citep{Zonca2019}. The spherical harmonic transforms can be converted to a flat-sky power spectrum via its approximation \citep[e.g.][]{khatri2016}. It is also possible to project a HEALpix map into flat-sky, though this projection may leave artifacts in the power spectrum \citep[e.g.][where the artifacts were on scales smaller than the instrument beam and ultimately inconsequential.]{romero2024}

        I do not endeavour to compare the accuracy of the various methodologies in this work. I will adopt the delta variance method as proposed in \citet{arevalo2012}, hereafter A12, as they validated its use in arbitrarily masked maps and showed improved accuracy over the delta variance method proposed in \citet{ossenkopf2008}. As advised in \citet{koch2019}, it is best to perform one's own validation tests to see how well a method performs in a particular scenario. 

        \subsubsection{Biases within A12 Delta Variance Method}
        \label{sec:A12_biases}
    
            The foundation of delta-variance approaches lie in the relation between the variance in the real domain and the Fourier domain. In the case of the Mexican hat filter proposed in A12, they note two equations for the variance for a filtering at a spatial frequency of $k_r$, denoted $V_{k_r}$ in their appendix. Specifically, they write for a general $n$-dimensional case:
            \begin{align*}
                V_{k_r} &= \int (F * I)^2 d^n x \\
                        &= \int P(k) | \hat{F}_{k_r} (k)|^2 d^n k,
                        \label{eqn:Variance_A12}
            \end{align*}
            where $I$ is the image, $F$ is the Mexican hat filter, with $F * I$ denoting the convolution of the two terms and $x$ denoting spatial dimensions; $P(k)$ is the power spectrum of the image, and $\hat{F}$ is the Fourier transform of $F$ and $k$ denoting spatial frequencies. As $P(k)$ is the quantity of interest, the method in A12 makes the approximation that this term can come out of the integral. While this makes this method relatively easy to use, the drawback is that this approximation can induce biases. For the case of a power-law spectrum with normalization $P_0$ and spectral index $\alpha$:
            \begin{equation}
                P_{\text{inj}} = P_0 k^{-\alpha},
                \label{eqn:spectral_pl}
            \end{equation}
            \cite{arevalo2012} find that there will be a normalization bias in the recovered spectrum $P_{\text{meas}}$: 
            \begin{equation}
                \frac{P_{\text{meas}}}{P_{\text{inj}}} = 2^{\alpha/2} \frac{\Gamma(n/2 + 2 - \alpha/2)}{\Gamma(n/2 + 2)}.
                \label{eqn:norm_bias}
            \end{equation}

            Of course, the underlying power spectrum may not be a single power-law, and there have been two additional expressions derived for additional Gaussian filtering. In particular, \citet{Zhuravleva2015} derive an expression for the influence when one subtracts Gaussian patches from an image, and \citet{romero2023} derive an expression for the bias imparted by a point spread function that can be characterized as a sum of Gaussians. 

        \begin{figure}[!h]
            \begin{center}
                \includegraphics[width=0.45\textwidth]{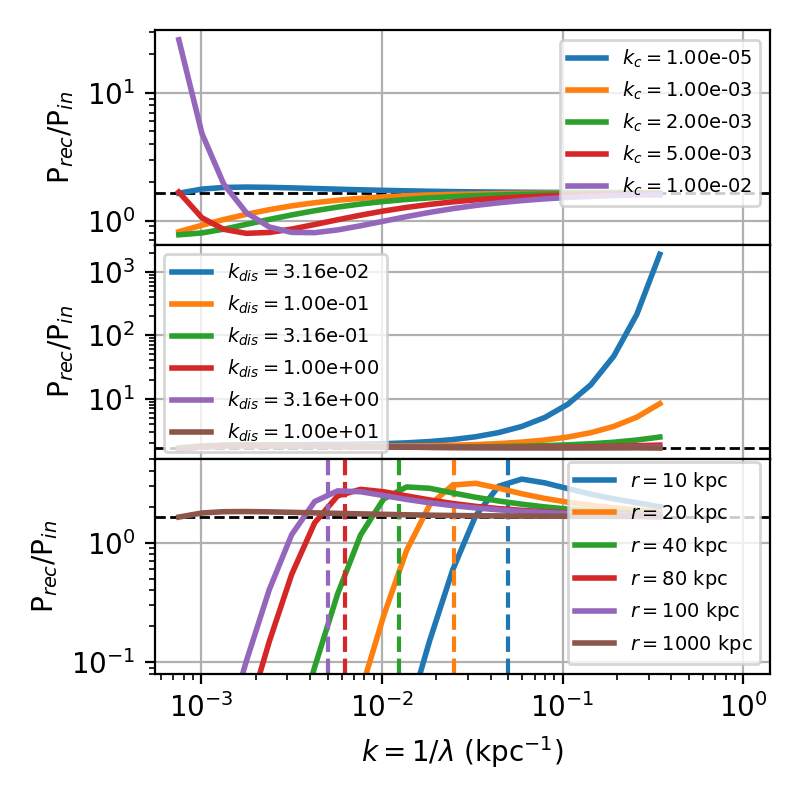}
            \end{center}
            \caption{Several biases to be aware of for the A12 method.  \textbf{Top:} Biases due to the spectral shape, notably the spectral turnover. Here we have adopted $\eta_c = 1$ and $k_{\text{dis}} \gg 1$. \textbf{Middle:} Biases due to the spectral shape, but regarding an exponential cutoff at high $k$. In particular, we have set $k_c = 1e-5$ and vary $k_{\text{dis}}$; $\eta_d = 1$. \textbf{Bottom:} an example of a masking bias. We adopt $k_c$ and $k_{\rm dis}$ such that our input spectrum is effectively a pure power law with known normalization bias denoted by the horizontal black dashed line. The legend denotes the radius of the circular mask; the vertical dashed lines indicate the wavenumber corresponding to the diameter of the mask.}
        \label{fig:A12_biases}
        \end{figure}

            Speaking more generally, we see that the magnitude and shape of the bias being induced depends on the \textbf{shape} of the underlying spectrum. The top and middle panels in Figure~\ref{fig:A12_biases} show how the biases behave for some example spectra, notably different $k_c$ and $k_{\text{dis}}$ for $\eta_c = 1$ and $\eta_d = 1$. Note that the bias at the dissipation end is more significant as the assumed slope is already steep ($\alpha = 11/3$); that is, from the normalization bias, we could expect any further steeping to induce larger biases, as happens. Conversely, at the cutoff end, the power spectrum must first pass through its peak (slope equal to zero), where the normalization bias is small, before the slope becomes steeply positive and incurs strong biases. 

            %\begin{multline}
            %    \frac{\Tilde{P}}{P} (k_r) = \left[ \sum_{i=1}^N \sum_{j=1}^N c_i c_j \left( \frac{2 x_i^2 x_j^2 + x_i^2 + x_j^2}{2  x_i^2 x_j^2} \right)^{-(n/2 + 2 - \alpha/2)} \right] \\\times \frac{\Gamma(n/2 + 2 - \alpha/2)}{\Gamma(n/2 + 2)} 2^{\alpha/2} .
            %    \label{eqn:PSFbias}
            %\end{multline}
            \begin{multline}
                \frac{\Tilde{P}}{P} (k_r) = \frac{\Gamma(n/2 + 2 - \alpha/2)}{\Gamma(n/2 + 2)} 2^{\alpha/2} \times \\ \left[ \sum_{i=1}^N \sum_{j=1}^N c_i c_j \left( \frac{2 x_i^2 x_j^2 + x_i^2 + x_j^2}{2  x_i^2 x_j^2} \right)^{-(n/2 + 2 - \alpha/2)} \right].
                \label{eqn:PSFbias}
            \end{multline}
            
            The adopted mask will induce its own bias; thus termed a \textbf{mask} bias. This \textbf{mask} bias has a spectral dependency too; therefore, to emphasize the bias is due to the mask, I choose parameters for our power spectrum model (Equation~\ref{eqn:ps_model}) such that I effectively model a pure power law. I then measure the power spectrum in various circles (about the center of the image). The bottom panel of Figure~\ref{fig:A12_biases} shows the mask biases incurred. In particular, there is decent power recovery up until scales corresponding to the diameter of the circular masks (vertical dashed lines), at which point the recovery falls off dramatically, which coincides with scales where one has few samples and sample variance (see Section~\ref{sec:uncertainties}) will dominate.
            I note that there is also a gradual overestimation in power from small scales up until the diameter of the mask. 
            %In all panels of Figure~\ref{fig:A12_biases}, the horizontal black dashed line corresponds to the known normalization bias for the assumed power-law index $\alpha = 11/3$.

            %While I have aimed to highlight the bias due to the mask itself here, the mask bias does depend on the spectral shape as well. For synthetic spectra generated in 2D one can numerically calculate the total bias and divide by the shape bias to arrive at the mask bias.

    \subsection{A note on (de)projection and its associated bias}

        A so-called projection bias is induced due to the approximation (Equation~\ref{eqn:formalism_approx}) of Equation~\ref{eqn:formalism_full}. Given that the projection dependence on $W(\theta,z)$ (in Fourier space, $|\tilde{W}(\theta,k_z)|$) reveals a dependence on the projected radius, this needs to be taken into account. In particular, the projection bias for spectra measured over a given area can be calculated as the area-averaged projection bias. 
        
        By extension, the variable $N(\theta)$ (Equation~\ref{eqn:NofTheta}) will have a corresponding area-averaged (effective) value, deemed $N_{\text{eff}}$. Alternatively, one can produce a map of $\sqrt{N(\theta)}$ and divide a $\delta y / \bar{y}$ or $\delta S / \bar{S}$ by this map. In doing so, the measured spectrum, while being measured in 2D space will correspond to the deprojected (3D) spectrum. For clarity, I explicitly label these power (or amplitude) spectra with $\delta y / \bar{y} / \sqrt{N}$ or $\delta S / \bar{S} / \sqrt{N}$ in the subscript.

    \subsection{Recovery of single underlying power spectrum}
    \label{sec:singleP3Drecovery}

        To perform a numerical check on the projection bias over a large region of a cluster, I inject pressure fluctuations, $\delta P / \bar{P}$, described by a single power spectrum ($P_{\text{3D}}$). For a given pressure profile model, $\bar{P}$, I integrate the quantity $\bar{P} * (1 + \delta P/\bar{P})$ along the line of sight to produce a Compton $y$ map. Here, I take $\bar{P}$ to be a spherical pressure distribution described by the A10 \citep{arnaud2010}\footnote{This may also be refered to as a UPP model. The parameters are those derived from the full sample in \citet{arnaud2010}.} pressure profile for a cluster at $z=0.3$ with $M_{500} = 6 \times 10^{14} M_{\odot}$. An integration of the UPP model without fluctuations yields a $\bar{y}$ map. From the two maps I produce a $\delta y/\bar{y}$ map, on which power spectra are measured, via FFT (no masking; over the entire image) and via A12 out to $R_{500}$.

        \begin{figure}[!h]
            \begin{center}
                \includegraphics[width=0.45\textwidth]{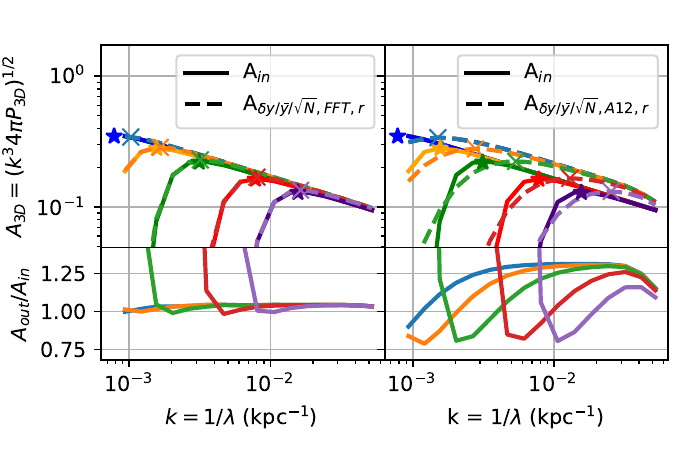}
                \includegraphics[width=0.45\textwidth]{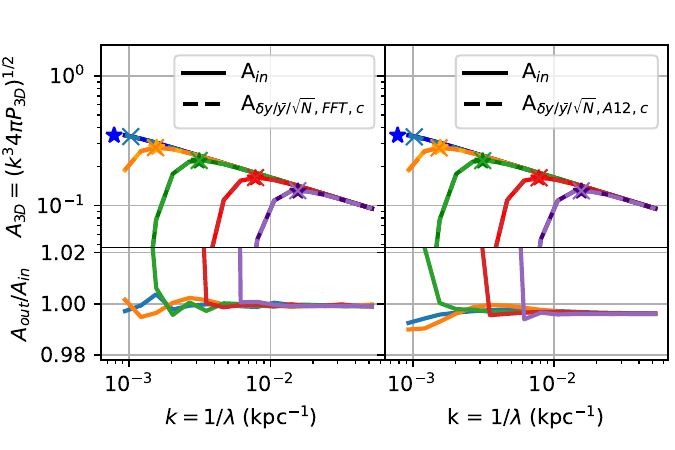}
            \end{center}
            \caption{Dividing by a map of $\sqrt{N}$ uncovers a well-normalized image with the underlying $P_{3D}$ power spectrum. The recovered spectra suffer from distortions due to projection, but otherwise this is good.}
        \label{fig:singleP3D_mod}
        \end{figure}
        
        Here, I am primarily concerned with the fidelity at large scales. As such, I adopt the model parameterization (Equation~\ref{eqn:ps_model}) with five different values of $k_c$: [5e-4, 1e-3, 2e-3, 5e-3, 1e-2] kpc$^{-1}$, all with $\alpha=11/3$, $\eta_{\rm H} = 4$, $P_0 = 1e{-4} {~\rm kpc}^{3}$, and I ignore the dissipation term (in practice $k_{\text{dis}} = 1e3$ kpc$^{-1}$). For each set of parameters, 100 realizations are created of which power spectra are calculated. The recovered power spectra at taken as the average across these realizations for each set of parameters.        

        Figure~\ref{fig:singleP3D_mod} shows the recovered amplitude spectra on images of $\delta y / \bar{y} / \sqrt{N}$ using both the FFT and A12 methods. The top panel shows the uncorrected amplitude spectra, while the bottom panel shows the corrected spectra. 
        %Note that the range of the ordinate of the fractional recovery ($A_{\text{out}}/A_{\text{in}}$) plots changes between the top and bottom panels. That is, at first glance, the recovery may look better for the uncorrected FFT spectra, but this is not the case. In all cases, the inferred spectra begin to differ dramatically below their respective cutoff wavenumber, $k_c$. 
        The corrected spectra are in excellent agreement with the input spectra, and only at extremely steep slopes does numerical resolution reveal a limitation of this check.

        %I further interrogate our results by calculating the inferred injection scales, i.e. peaks of the amplitude spectra with high spectral sampling. I find that in the uncorrected spectra, the FFT method will recover $k_{\text{inj}}$ roughly 4\% above the true value, while the A12 method recovers values roughly 70\% above the true value. However, the uncorrected spectra are distorted such that the recovery of the peaks is reasonable; the FFT method recovers the peak (of the amplitude spectra) within 3\%, and the A12 method recovers the peak within 1\% of the true value. When I correct for the spectral biases, both the FFT and A12 methods recover $k_{inj}$ to better than 1\%; however, the amplitudes are biased high by just over 1\%.

        Although the shape of spectra explored here is rather uniform, it may be of interest to consider some recovered quantities before and after correcting for biases. Excluding the injected spectrum with $k_c = 5e-4$, the uncorrected spectra yield $k_{\text{inj}}$ ($A(k_{\text{inj}}$) 4\% (3\%) and 70\% (1\%) higher for the FFT and A12 methods, respectively. For the corrected spectra, the FFT and A12 methods recover both $k_{\text{inj}}$ and $A(k_{\text{inj}}$ within 1\%. I find that these results are nearly identical if one calculates the spectra on $\delta y / \bar{y}$ images and subsequently divides by $N_{\text{eff}}$.  

    \subsection{Which residual map to use}
    \label{sec:smoothing}

        %It is well established that one must account for the power spectrum of an instrument's beam (or point-spread-function, PSF) when analyzing the power spectra of ICM fluctuations. However, many treatments resolve the matter be effectively assuming that $I_{\text{rel}}$ is convolved with the point spread function, such that one is performing Fourier analyses on the image 
        %$I_{\text{rel}}^{\prime} = I_{\text{rel}} * {\rm PSF}$. In practice, I don't have direct access to $\delta S$ (and consequently I neither have access to $I_{\text{rel}}$), but rather $\delta S^{\prime} = \delta S * {\rm PSF}$. Moreover, deconvolution of images tends to increase the noise. However, because simple models of the ICM surface brightness can leverage the reduction in noise over many beams, it is generally quite feasible to infer a deconvolved ("unconvolved") ICM surface brightness model, $S_{\text{ICM}}$. It should be clear that a beam-convolved model, $S_{\text{ICM}}^{\prime}$ is also easily attained, either by convolving $S_{\text{ICM}}$ by the beam, or simply by fitting a model directly to an image without performing any deconvolution.

        The formalism presented in the introduction and checked in just the previous section is calculated in the case that there is no beam convolution. The canonical correction to beam convolution is to simply divide the inferred power spectrum (after noise debiasing) by the power spectrum of the beam. However, this correction is only approximate as it is not the map of $\delta I/\bar{I}$ that is convolved by the beam. Rather, $I$ is smoothed by the beam, call this $I^{\prime}$, and the residual can be labeled $\delta I^{\prime}$. I can consider the analogous model variables $\bar{I}$ and $\bar{I}^{\prime}$.

        Noting that I do not have access to $\delta I/\bar{I}$, one may ask: should I prefer $\delta I^{\prime}/\bar{I}$ or $\delta I^\prime/\bar{I}^{\prime}$? An implied question is how accurately do these fractional residuals recover the true spectrum?

    %\textcolor{red}{As is the theme for this paper, in order to provide some more quantitative response to this question, I must make some choices which benefit from motivations presented later in the paper.} 
        To address this question I build on the infrastructure previously used to generate pressure fluctuations and generate $y$ and $\bar{y}$ maps. I adopt Gaussian beams with full-width half-maxima (FWHM) of $5^{\prime\prime}$, $10^{\prime\prime}$, and $20^{\prime\prime}$ to generate corresponding $y^{\prime}$ and $\bar{y}^{\prime}$ maps. I adopt a single power spectrum of fluctuations with $k_c = 1e-3$ kpc$^{-1}$ and the remaining parameters as set previously.

        As the importance of the smoothing should depend on the angular size of the cluster relative to the beam, I explore a range of cluster sizes. Specifically, I choose to sample a grid of clusters with masses $M_{500} = [2,4,6] \times 10^{14} {\rm M}_{\odot}$ at redshifts $z = [0.1,0.3,0.5,0.7,0.9]$. For the assumed cosmology, this translates to $1.6^{\prime} < R_{500} < 12.4^{\prime}$. 
        %As before, I opt to model toy SZ fluctuations due to the convenience of a "universal pressure profile", wherein I adopt the A10 model as before. For each (M$_{500}$,$z$) I generate a 2D image with power spectra as in Equation~\ref{eqn:ps_model}, $k_{\rm c} = 1e{-3} {~\rm kpc}^{-1}$, $\alpha = -11/3$, $k_{\rm dis} = 1e1 {~\rm kpc}^{-1}$, and $P_0 = 1e{-2} {~\rm kpc}^{3}$ and scale pixels by $\sqrt{N(\theta)}$. This image thus constitutes $I_{\text{rel}}$, which I will equate to $\delta y/\bar{y}$ here given the specific application to the SZ surface brightness fluctuations. By multiplying the $\delta y/\bar{y}$ map by a map of $y$, I obtain a $\delta y$ map. Subsequently I can assume a beam and obtain beam-smoothed maps, $\delta y^{\prime}$ and $y^{\prime}$. From these, I can obtain $\delta y^{\prime}/\bar{y}$ and $\delta y^{\prime}/\bar{y}^{\prime}$ images. For these clusters, I assume beams which are symmetric Gaussians with full-width half-maxima (FWHM) of $5^{\prime\prime}$, $10^{\prime\prime}$, and $20^{\prime\prime}$.

        Now, I simply need to compare the power spectra of these images. However, as indicated in Section~\ref{sec:PS_estimator}, measuring a power spectrum is not necessarily trivial. There is also the matter of the region within which to measure the power spectrum, discussed further in Section~\ref{sec:synthesis}. As will be motivated in Section~\ref{sec:synthesis}, I calculate the power spectra in three regions: $\theta < 0.4 R_{500}$, $0.4 R_{500} < \theta < 1.0 R_{500}$, and $1.0 R_{500} < \theta < 1.5 R_{500}$. I employ the delta-variance method detailed in \citet{arevalo2012} and account for the induced PSF bias explained in Section~\ref{sec:A12_biases}.

\begin{figure}[!h]
            \begin{center}
                \includegraphics[width=0.48\textwidth]{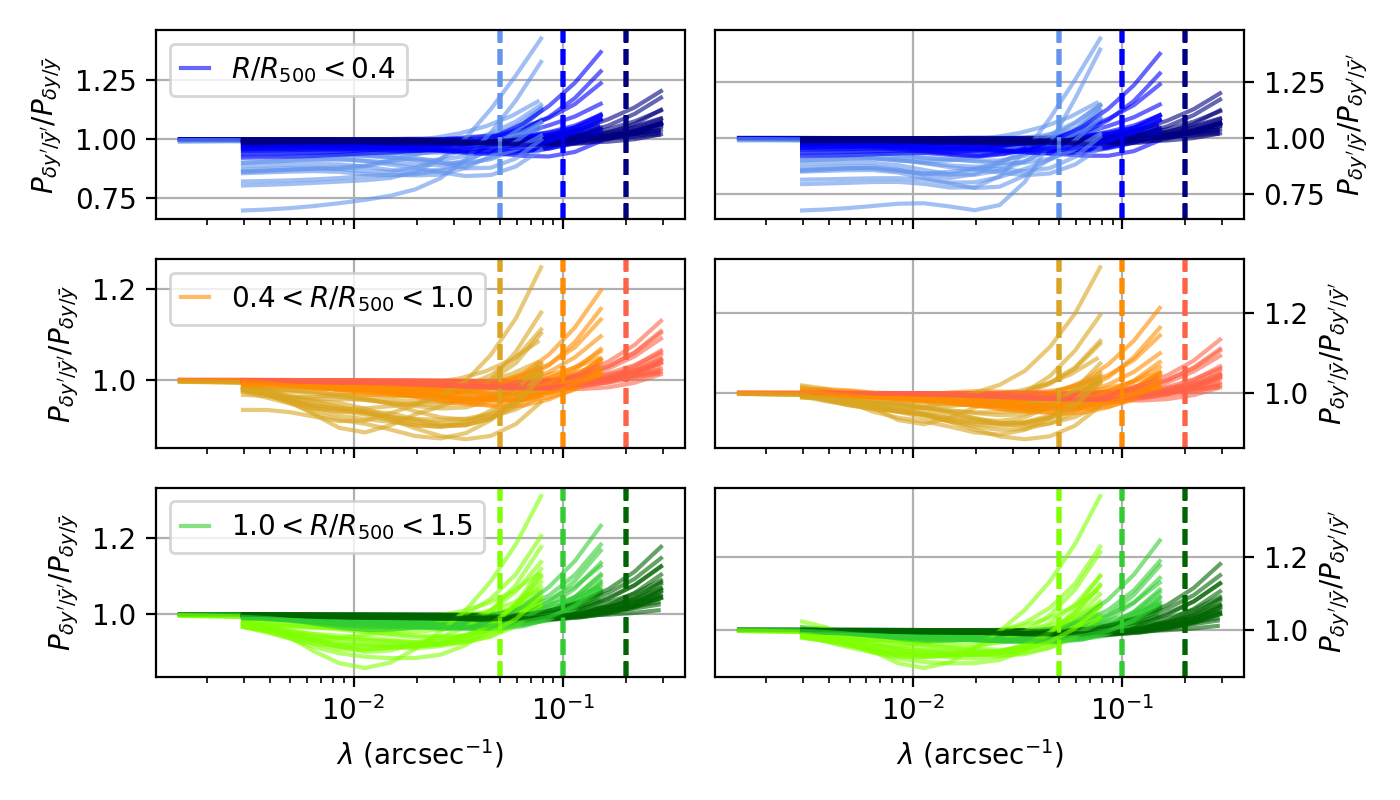}
            \end{center}
            \caption{The left column compares the recovered $P_{\delta y^{\prime}/\bar{y}^{\prime}}$ relative to $P_{\delta y/\bar{y}}$, while the right column compares $P_{\delta y^{\prime}/\bar{y}}$ relative to $P_{\delta y/\bar{y}}$. Both $P_{\delta y^{\prime}/\bar{y}^{\prime}}$ and $P_{\delta y^{\prime}/\bar{y}}$ have been corrected for canonical PSF bias via Equation~\ref{eqn:PSFbias}.
            The shapes of of the relative curves in the two columns are very similar. For each region, the darkest shade corresponds to smoothing by a beam with FWHM = $5^{\prime\prime}$, and the lightest shade as a smoothing with FWHM = $20^{\prime\prime}$. }
            % This plot is rather insensitive to the assumed underlying power spectrum
        \label{fig:P2D_vs_smoothing}
        \end{figure}

        Figure~\ref{fig:P2D_vs_smoothing} shows the recovered power spectra of either the $\delta y^{\prime}/\bar{y}^{\prime}$ image relative to the power spectra of $\delta y/\bar{y}$ (left column) or the power spectra of $\delta y^{\prime}/\bar{y}$ relative to $\delta y^{\prime}/\bar{y}^{\prime}$ (right column). To reiterate, the (standard) PSF corrections (discussed in Section~\ref{sec:A12_biases}) are already accounted for in $P_{\delta y^{\prime}/\bar{y}}$ and $P_{\delta y^{\prime}/\bar{y}^{\prime}}$ and thus the motivation for this investigation. The column on the right indicates that there is not a significant difference between taking power spectra of $\delta y^{\prime}/\bar{y}^{\prime}$ versus $\delta y^{\prime}/\bar{y}$.
        
        I define biases, $b_{y^{\prime}/\bar{y}}$ as $1 - P_{\delta y^{\prime}/\bar{y}} / P_{\delta y/\bar{y}}$ if $P_{\delta y^{\prime}/\bar{y}} / P_{\delta y/\bar{y}} > 1$ otherwise as $1 - P_{\delta y/\bar{y}} / P_{\delta y^{\prime}/\bar{y}}$. I define $b_{y^{\prime}/\bar{y}^{\prime}}$ analogously. Secondarily, I calculate the ratio between the appropriate FWHM and $\theta_{500}$, the angular size of $R_{500}$. I find that the maxima of both $b_{y^{\prime}/\bar{y}}$ and $b_{y^{\prime}/\bar{y}^{\prime}}$ are well approximated as $\text{FWHM} / \theta_{500}$.

        \subsubsection{Remarks on smoothing}
        \label{sec:smooth_remarks}

            I find no clear preference between the use of $\delta y^{\prime}/\bar{y}$ and $\delta y^{\prime}/\bar{y}^{\prime}$ images. It is possible that different region selections and assumed underlying spectra change the outcome; though in the latter case I did not discern a strong dependence on the underlying spectrum for the range I investigated. This bias may also change for X-ray analyses given the steeper drop in surface brightness profile. However, given the comparatively small PSF of current and future X-ray instruments, I do not perform an analogous investigation.
            
            While this bias can ostensibly be corrected if the underlying spectral shape is known, the the scaling of the bias indicates that situation (FWHM relative to $\theta_{500}$) where the bias is greatest is where the constraints on the the spectral shape will be worst as the scales accessible will be most limited. This suggests that in practice this bias will be quite difficult to combat. If one adopts target uncertainties near $\lesssim 10$\%, then this investigation suggest that fluctuations studies are best suited to cases where $\theta_{500} \gtrsim 10$FWHM.
            %Given the nearly linear relation between a maximal bias and the beam size relative to the $\theta_{500}$, if one is 

    \subsection{Multiple underlying spectra}
    \label{sec:multiple_spectra}

        The evidence that there is not a single underlying $P_{3D}$ within the ICM for any thermodynamic quantity is substantial. Of those studies of surface brightness fluctuations that have investigated multiple regions (or apertures about a center) have found that the inferred $P_{3D}$ vary with region or aperture size \citep[e.g.][]{Zhuravleva2015,khatri2016,romero2023,dupourque2023}. This is not only expected, but it is in fact a requirement if turbulent non-thermal pressure increases with cluster-centric radius \citep[e.g][]{battaglia2012a,nelson2014a,biffi16,angelinelli2020}.
        
        \subsubsection{A plausible scenario}
        \label{sec:plausible}
            
            \begin{figure}
                \begin{center}
                    \includegraphics[width=0.45\textwidth]{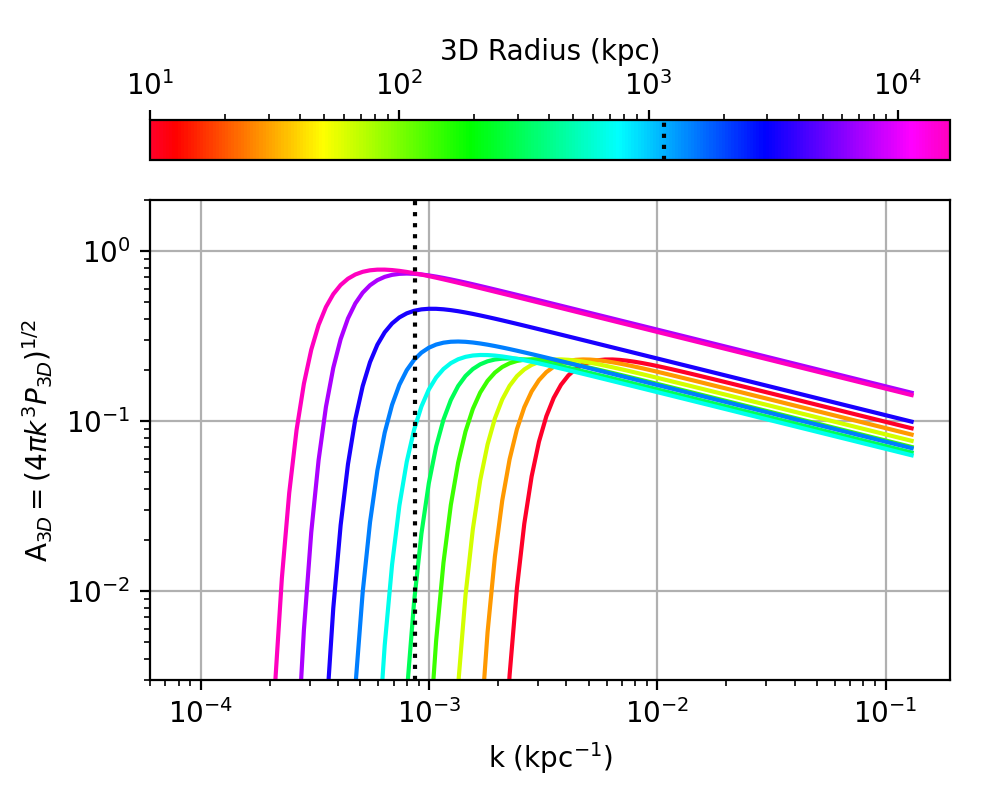}
                \end{center}
                \caption{The resultant $A_{3D}$ spectra as measured by the A12 method relative to the volume-weighted $A_{3D}$ per shell.}
            \label{fig:PlausibleP3Ddistribution}
            \end{figure}

            In this section I explore the noiseless effects on projected spectra when there exists a distribution of power spectra throughout a cluster. In addition to the above notes on biases and spectral weighting, Appendix~\ref{sec:multP3D_formalism} revisits the formalism of a projected power spectrum in the case of multiple underlying $P_{\text{3D}}$.
            %As noted in \citet{romero2023} and evidenced in Figure~\ref{fig:multipleP3D_numerical_int}, multiple power spectra can potentially flatten the resultant peak of the power spectrum in a region. If one is measuring power spectra in radial (annular) bins, either an onion-layered deprojection or forward modelling may be considered as a means to accurately recovering the 3D power spectrum of corresponding 3D radii to the projected radii of a given annulus. 
            
            \begin{figure*}
                \begin{center}
                    \includegraphics[width=0.45\textwidth]{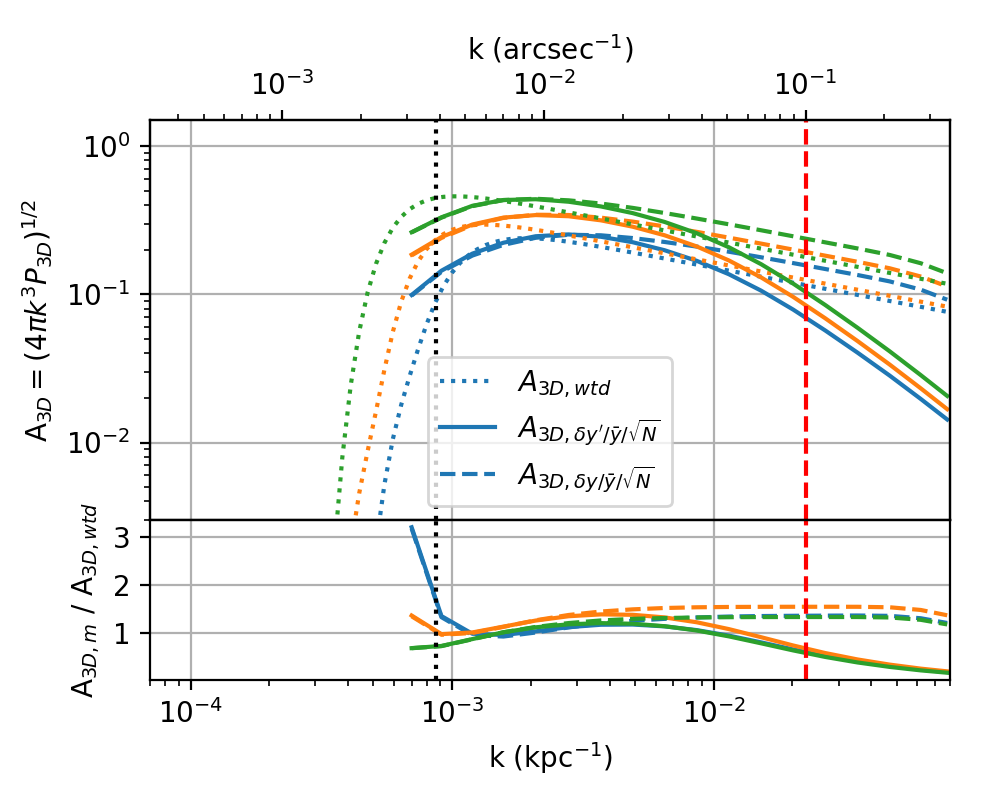}
                    \includegraphics[width=0.45\textwidth]{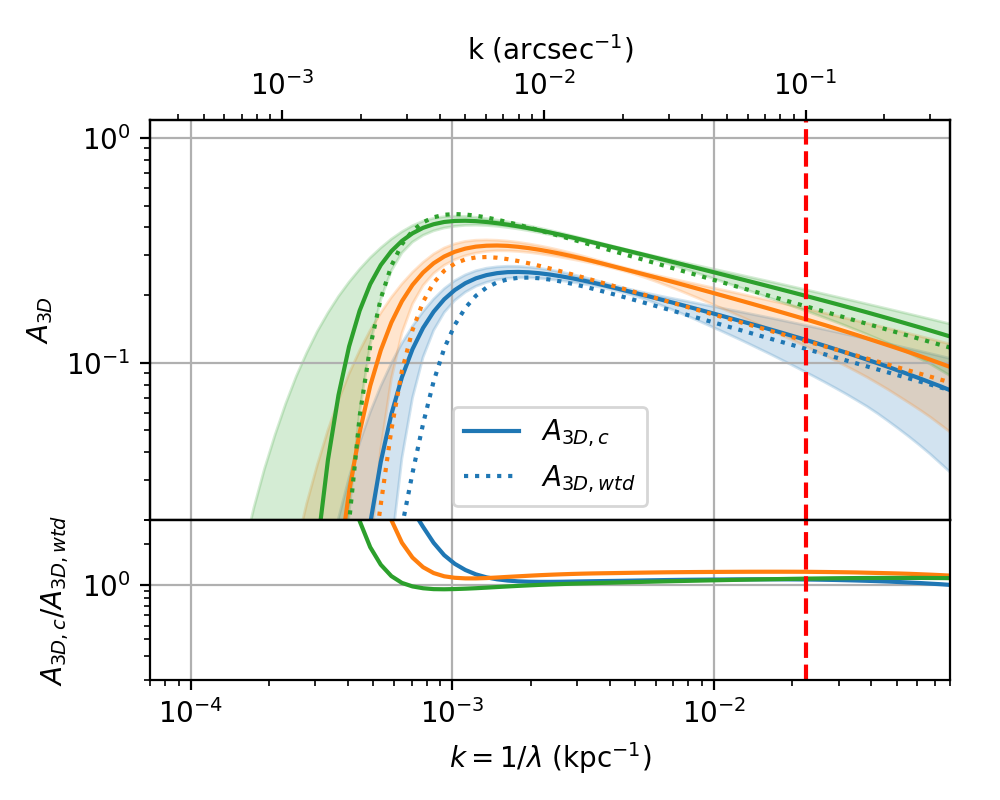}
                \end{center}
                \caption{\textbf{Left:} The amplitude spectra as recovered via the A12 method with no bias corrections applied. $A_{\text{3D,wtd}}$ refers to the pressure-weighted average of the input spectra (within the corresponding 3D shell); shown as dotted curves. The solid curves are the recovered spectra when assuming observations with an instrument that has a Gaussian FWHM of $10^{\prime\prime}$, while the dashed curves are the recovered spectra with no PSF distortion. \textbf{Right:} Best fit amplitude spectra to the raw spectra as observed with an instrument with FWHM of $10^{\prime\prime}$, with bias correction included. Fits are performed only with measured points at scales larger than the FWHM of $10^{\prime\prime}$ (indicated by the dashed red line). Shaded regions indicate $1\sigma$ intervals about the best fit. In all panels, blue, orange, and green curves correspond to Ring 1, Ring 2, and Ring 3, respectively.}
            \label{fig:RecoveredP3D}
            \end{figure*}

            %To investigate how one can accurately recover the underlying, pressure-weighted, $P_{\text{3D}}$ within a corresponding volume, 
            I adopt a schema wherein I have ten different $P_{\text{3D}}$ with $\alpha=11/3$ and no dissipation term. The ten power spectra occupy spherical shells (with tapered transitions) and a peak chosen such that the non-thermal pressure follows the radial profile found in \citet{nelson2014a}; see Section~\ref{sec:approach} for a more details on the non-thermal pressure profile. As the injection scale is expected to increase with cluster-centric radius \citep[$r$; e.g.][]{dupourque2023,romero2024}, I assume that $l_{\rm inj} \propto r$. Once again, I assume an A10 profile of a $z=0.3$ and $M_{500} = 6 \times 10^{14} M_{\odot}$ cluster. The input power spectra are shown as amplitude spectra in Figure~\ref{fig:PlausibleP3Ddistribution}.
            
            As was done in Section~\ref{sec:smoothing}, I adopt regions (rings) with edges of [[0.4,1.0,1.5] $R_{500}$ (see also Section~\ref{sec:synthesis}). Spectra from maps of $\delta y / \bar{y} / \sqrt{N}$ and $\delta y^{\prime} / \bar{y} / \sqrt{N}$ are calculated. In the left panel of Figure~\ref{fig:RecoveredP3D} are raw (measured) amplitude spectra insofar as they are the values recovered by the A12 method without any bias corrections applied and averaged over realizations. If considered from the perspective of not knowing the input spectra, we may hope to fit a parameterized power spectrum model to our recovered spectra, calculate biases, and infer bias-corrected spectra. This should be done iteratively until convergence. 
            
            It is computationally easy to calculate bias corrections for spectral shape, but more expensive to correct for a mask bias. That is, for each fit performed, a bias should be calculated. In practice, I use \lstinline{scipy.optimize.curve_fit} to fit for parameters in Equation~\ref{eqn:ps_model}, where I put strong priors on $\alpha$, $\eta_c$, and $\eta_d$ and include shape bias corrections with each spectral model. After each iteration of \lstinline{scipy.optimize.curve_fit}, I calculate a mask correction for the obtained power spectrum. I repeat these two steps until convergence. While I am working with noiseless synthetic data, the standard deviation of the power spectra across realizations imparts sample variance, which is used in the fitting procedures.

            \begin{deluxetable}{ccccc}[!h] 
    \centering 
    \tablecaption{Peaks of Amplitude Spectra\label{tbl:as_peaks}} 
    \tablehead{ 
     \colhead{ \hspace{0.05cm}Region\hspace{0.05cm} } &\colhead{ \hspace{0.5cm} \hspace{0.5cm} } & \colhead{ $\ell_{\text{inj}}$ (kpc) } & \colhead{ \hspace{0.05cm}$A_{\text{3D}}(k_{\text{peak}})$\hspace{0.05cm} } & \colhead{ \hspace{0.05cm}$\sigma_{\text{3D}}$\hspace{0.05cm} }
      } 
    \startdata 
    \multirow{4}{*}{Ring1} & Expected & 515     & 0.24 & 0.33 \\
     & Raw (10) & 359 & 0.25 & 0.35 \\
     & Raw  & 359 & 0.25 & 0.38 \\
     & Corrected & 559 & 0.25 & 0.37 \\
    \hline 
    \multirow{4}{*}{Ring2} & Expected & 746     & 0.30 & 0.40 \\
     & Raw (10) & 479 & 0.34 & 0.47 \\
     & Raw  & 479 & 0.34 & 0.51 \\
     & Corrected & 673 & 0.33 & 0.47 \\
    \hline 
    \multirow{4}{*}{Ring3} & Expected & 1003     & 0.46 & 0.58 \\
     & Raw (10) & 479 & 0.44 & 0.60 \\
     & Raw  & 479 & 0.44 & 0.65 \\
     & Corrected & 890 & 0.43 & 0.58 \\
    \enddata 
    \tablecomments{Peaks of A$_{\text{3D}}$.}
    \vspace{-1cm} 
\end{deluxetable}

            In this example, where the cutoff scale is within the measured range, I find that the spectral shape bias is dominant, though the mask bias is still important as it is not unity. Given the convergence from 
            \lstinline{scipy.optimize.curve_fit}, I fix the mask bias to its converged value. I subsequently run spectral fits again, but through \lstinline{emcee} over 10000 steps without priors on spectral parameters. From the MCMC chains I recover amplitude spectra shown in the right panel of Figure~\ref{fig:RecoveredP3D}.
            %In this example, I find that the mask+shape bias is very similar to the shape bias itself. I divide the mask+shape bias by the shape bias to calculate the what the bias contribution is from the mask itself and I find that this contribution is quite small compared to the shape bias, but it is not unity. I take this as indication that it may be sufficient to fix the mask bias to its converged value above and to revisit fitting for all the parameters (without priors) using \lstinline{emcee} using 10000 steps.   

            These (bias-corrected) fits show amplitude spectra that are much better matched to the expected amplitude spectra than the raw spectra are matched to the expected spectra. 
            %Even so, I might have expected even greater agreement for a scenario where the only uncertainty is the sample variance. 
            The enlarged uncertainty envelopes (shaded regions) at high $k$ is due to the degeneracy between the the PSF bias correction and the slope $\alpha$. Conversely, at low $k$, there is also a degeneracy between the (shape) bias correction and the rate of exponential cutoff ($k_c$ and $\eta_c$). When including measurement errors, it is likely that use of priors will be warranted (especially for terms like $\eta_c$.
            %While fitting for $k_{\text{dis}}$ and $\eta_{\rm d}$ is not warranted for the input dissipation scale chosen ($k_{\text{dis}} = 1e3$ kpc$^{-1}$), this is not the primary cause for the large shaded envelopes towards higher $k$; rather, the PSF, cascading slope $\alpha$ and their combined bias correction have a degeneracy. Similarly, the cutoff and its bias correction have a degeneracy. Assuming a fixed value for $\eta_{\rm c}$ (and omitting a dissipation term) may help.
            %but the recovered peaks are still below the expected peaks. 

            I investigated incorporating projection biases and found that attempting to correct for projection biases assuming the recovered (corrected) spectrum within a ring generally produced worse results. The problem is that a proper calculation of the projection bias requires knowing the detailed distribution of power spectra (cf Appendix~\ref{sec:multP3D_formalism}). 
            %If $P_{\text{3D}}(r)$ is continually evolving with radius, a single $P_{\text{3D}}$ within a sphere or shell may easily be insufficient to account for projection bias. Said another way, one is more likely to miscalculate the projection bias than to calculate it correctly.
            
            %From the previous section (Section~\ref{sec:multiple_spectra}) one recalls that the transition between different power spectra is important, whether one attempts to model the distribution of power spectra as being discrete with some taper, or more physically, a smooth transition through a range of power spectra. The point is, without quite detailed knowledge of this distribution, 
            
            Additionally, I investigated onion peeling the power spectra and found that this does improve the recovered spectrum. For brevity, I present the findings of onion peeling in an investigation of power spectrum recovery with added measurement noise in Appendix~\ref{sec:ShapeRecovery}. In the scenario here, $A_{\text{3D},\delta y^{\prime}/\bar{y} / \sqrt{N}}$ matches $A_{\text{3D},\delta y/\bar{y} / \sqrt{N}}$ at low to moderate $k$, reinforcing the inference from Section~\ref{sec:smoothing} that for this cluster ($z = 0.3$, $M_{500} = 6\times 10^{14} \text{M}_{\odot}$), the (secondary) bias smoothing due to a $10^{\prime\prime}$ beam will be small.

            %\textcolor{red}{OK, I discussed some difficulty of fitting the spectra (i.e. there are still uncertainties). What I need to still address are (1) How well they are recovered (or not), and (2) mention that attempting an onion-peeling of sorts is probably not warranted in most cases.}
            %Few of the parameters in Equation~\ref{eqn:ps_model} are directly of interest: $\alpha$ being chief among them. The primary interest in their descriptive capacity with respect to the shape and amplitude of the power spectra. Because this study is not focused on high-$k$ values and attempting to access dissipation scales, the primary points of interest are (1) recovering the injection scale(s), $l_{\rm inj}$ and (2) inferring the turbulent Mach number(s). These two values are not dependent on a single parameter and can be estimated without Equation~\ref{eqn:ps_model} to the data (e.g. $l_{\rm inj}$ is simply $1/k_{\rm peak}$, where $k_{\rm peak}$ is the location of the peak in a given 3D amplitude spectrum.

            %To estimate the Mach number from amplitude or power spectra, there are currently two principle avenues. The first is to use a relation between the peak of a 3D amplitude spectrum \citep[][see also Section~\ref{sec:approach}]{Gaspari2013_PS} and the second approach is to calculate the logarithmic standard deviation in fluctuations via integrating spectra \citep[][]{zhuravleva2012,simonte2022,zhuravleva2023}. The two approaches have been found to generally be in agreement, especially when spectra are sufficiently well recovered \citep[e.g.][]{romero2023,dupourque2024,romero2024}.

            In the context of relating spectra of fluctuations to turbulent velocities, Table~\ref{tbl:as_peaks} reports on key spectral quantities for the input spectra and recovered spectra. As seen in Section~\ref{sec:singleP3Drecovery}, the raw amplitude spectra decently capture the amplitude of the peak, but it will underestimate the injection scale (in this scenario by $\sim50$\%).  Table~\ref{tbl:as_peaks} also includes the dispersion in fluctuations (taken as the square root of the integrated spectra).

    \subsection{Remarks}
    \label{sec:remarks}

        %In the case of these noiseless (aside from sample variance) investigations, I have taken pains to inventory biases in computing power spectra and have been able to validate that power spectra can be measured across a large region of a galaxy cluster without concern that the dependency of the Window function, $W(\theta,z)$, on $\theta$ will irreparably distort the projected (2D) power spectrum relative to an underlying 3D power spectrum of pressure or density fluctuations. The formalism equally allows for multiple power spectra of 3D (pressure or density) fluctuations throughout the cluster. 
        
        %When considering a plausible distribution of multiple power spectra, I find that I am able to recover, with a modicum of accuracy, the 3D power spectra in 3D regions corresponding to their projected 2D regions (annuli, in my scenario) without onion-peeling. I do, however, include corrections biases due to the spectral shapes and due to masking. 

        In my investigation, I have found that the most important biases for accurate spectral recover are the spectral shape and masking biases, which are induced by the A12 method. While exploring these biases for other methods is beyond the scope of this paper, such biases should clearly be checked for whichever method is adopted.

        In the plausible scenario, the radially evolving $P_{\text{3D}}(r)$ also produces a "contamination" bias, where $P_{\text{3D}}(r > \theta)$ contaminate spectra within an annulus out to $\theta$. This appears to be minimal, but for sufficiently sensitive observations, applying an onion peeling approach may be helpful (see Appendix~\ref{sec:ShapeRecovery}). 

        Additional biases noted in this section include the projection bias and smoothing bias. However, I found that, for my plausible scenario, attempts to correct for the projection bias were detrimental. 
        %Given the expectation for an evolving $P_{\text{3D}}(r)$, I would view attempts to correct for a projection bias of a single power spectrum with caution.
        Given the relatively small impact of the smoothing bias expected given $\theta_{500} = 4^{\prime}.3$ and a Gaussian beam of $10^{\prime\prime}$, no treatment is applied here.

\section{Uncertainties in power spectra}
\label{sec:uncertainties}

    Uncertainties in power spectra have long been calculated assuming either an underlying Gaussian or Poisson random field \citep[e.g][]{Tukey1956,blackman1958,hoyng1976}. In Section~\ref{sec:PS_estimator}, I adopted the A12 method of calculating power spectra. In \citet{arevalo2012}, they provide a formula assuming that the data correspond to a Gaussian random field. They find that $\sigma_{P}(k) \propto P(k)/\sqrt{N}$ for $N$ total data points. 
    
    Reframing the what is presented in \citet{arevalo2012}, I find:
    \begin{equation}
        \sigma_{P}(k) \simeq P(k) \sqrt{2/N_k},
        \label{eqn:PS_uncertainty_vs_N}
    \end{equation}
    where $N_k = \Omega \pi k^2$ sampling elements for an sampling area $\Omega$ at a given frequency $k$. Note that this $N_k$ corresponds to the number of Gaussian smoothing kernels (those used in the Mexican hat filter) fit within a region of area $\Omega$. This scaling matches that found in \citet{arevalo2012} and is found to match results from previous works (see Appendix~\ref{sec:VerifySpecUnc}).
    Note that this scaling implies that the uncertainty in the power spectrum will be steeper than the power spectrum by 1 for a two dimensional dataset, i.e. in images.  

    It is useful to write the total measured power spectrum, $P_{\rm tot,m}$ as the sum of the measured power spectrum due to the signal, $P_{\rm s,m}$ and the measured noise component $P_{\rm n,m}$:
    \begin{equation}
        P_{\rm tot,m}(k) = P_{\rm s,m}(k) + P_{\rm n,m}(k).
        \label{eqn:PS_signal_noise}
    \end{equation}

    We are interested in the calculation of the signal power spectra and its uncertainty. We do not have direct access to $P_{\rm s,m}(k)$, but access it through $P_{\rm s,m}(k) = P_{\rm tot,m}(k) - P_{\rm n,m}(k)$. Let us consider $P_{\rm s,e}(k)$ to be the expected signal power spectrum. From this, we can calculate the expected sample variance, $\sigma_{P_{\rm s,e}(k)}$.    
    The total uncertainty on the inferred signal spectrum is then given by:
    \begin{equation}
        \sigma_{P_{\rm s,m}(k)}^2 = \sigma_{P_{\rm n,m}(k)}^2 + \sigma_{P_{\rm s,e}(k)}^2.
        \label{eqn:PS_uncertainty_signal}
    \end{equation}

    \subsection{White noise approximation}
    \label{sec:wn_approx}

        Note that sample variance, $\sigma_{P_{\rm s,e}(k)}$, does not improve with observational sensitivity. However, $\sigma_{P_{\rm n,m}(k)}$ will improve (by definition) with the sensitivity of observations. I will focus on the uncertainty due to noise, $\sigma_{P_{\rm n,m}(k)}$.

        With X-ray observations, the noise is Poisson, which equates to white noise. In SZ observations, the noise can be pink, but often at the scales of interest one tends to be in a regime where the noise is sufficiently white \citep[e.g.][]{romero2023,romero2024}. For the case of white noise, $P_{n}(k) = \text{constant}$, in a 2D image Parseval's theorem gives us
        \begin{equation}
            P_{n}(k) \pi k_{\rm max}^2 = V,
            \label{eqn:parseval_2D}
        \end{equation}
        where $V$ is the variance in the image and $k_{\rm max}$ is the highest frequency being sampled. In practice, the highest frequency is set by the (binned) pixel size for X-ray observations and by the FWHM of the beam for SZ observations.

    \subsection{Uncertainties within a circular region}
    \label{sec:circular_regioning}

        Let us assume that an observational facility points at the center of a (spherical) cluster. The variance in the image for both X-ray and SZ facilities can be reasonably well approximates as a function of $\theta$, the projected cluster radius.
        For SZ observations, let us denote the variance and standard deviation in the observed images as $V_y$ and $\sigma_y$, respectively. The variance in the normalized residual ($\delta y/\bar{y}$) map is then give by:
        \begin{equation}
            V_{\delta y/ \bar{y}}(\theta) = \left( \frac{\sigma_{y}(\theta)}{\bar{y}(\theta)} \right)^2,
            \label{eqn:fluctuation_noise_variance}
        \end{equation}

        In order to measure a power spectrum, one needs to do so over an area. The desired quantity is then the area-averaged variance. Let us consider the case of a circular region about the cluster center. The variance within a circle of radius $\theta$ can be expressed as:
        \begin{equation}
            \langle V_{\delta y / \bar{y}} \rangle(\theta) = \frac{ \int_0^\theta V_{\delta y / \bar{y}} 2\pi \theta^{\prime} d\theta^{\prime}}{\pi \theta^2}.
            \label{eqn:circular_var}
        \end{equation}
        Consequently, from equations~\ref{eqn:PS_uncertainty_vs_N} and \ref{eqn:parseval_2D} the uncertainty due to noise in a residual spectrum will be given by:
        \begin{equation}
            \sigma_{P_{\text{n,m},\delta y/\bar{y}}}(k,\theta) = \frac{\langle V_{\delta y / \bar{y}} \rangle(\theta)}{\pi k_{\rm max}^2} \sqrt{\frac{2}{N_k(\theta)}}.
            \label{eqn:circular_spectral_uncertainty}
        \end{equation}
        If we note that the dependence on $k$ simply governs a normalization factor for $\sigma_{P_{\text{n,m},\delta y/\bar{y}}}(k,\theta)$, at a given $\theta$, we can arbitrarily choose a $k$ to investigate the behavior over $\theta$.

        In the case of X-ray observations, the variance is given simply by the expected counts, such that for a surface brightness rate model ($\bar{S}_{\rm tot}$ with units counts/s/area), one can write:
        \begin{equation}
            V_{\text{counts}} = \bar{S}_{\rm tot} E \Omega_{\text{pix}},
            \label{eqn:Var_Counts}
        \end{equation}
        where $E$ is the exposure map (in seconds), and $\Omega_{\text{pix}}$ is the pixel area. The variance in the rate map is given by $V_{\text{counts}}/( E \Omega_{\text{pix}})^2$ and the variance in the $\delta S / \bar{S}$ image is given by:
        \begin{equation}
            V_{\delta S/\bar{S}} = \frac{\bar{S}_{tot}}{E  \Omega_{\text{pix}} \bar{S}_{ICM}^2}.
            \label{eqn:average_xvar}
        \end{equation}
        %By analogy, Equation~\ref{eqn:circular_var} can be used to calculate $\langle V_{\delta S / \bar{S}} \rangle$, and also by analogy Equation~\ref{eqn:circular_spectral_uncertainty} will yield $\sigma_{P_{\rm n,m}(k)}(\theta)$ for the X-ray case
        %Equation~\ref{eqn:circular_spectral_uncertainty} 
        Having arrived at an expression for $V_{\delta S/\bar{S}}$, the calculation of $\langle V_{\delta S / \bar{S}} \rangle$ and $\sigma_{P_{\text{n,m},\delta S/\bar{S}}}(\theta)$ follow as in Equations~\ref{eqn:circular_var} and \ref{eqn:circular_spectral_uncertainty} respectively. % where one simply replaces $V_{\delta y / \bar{y}}$ with $V_{\delta S/\bar{S}}$.

\begin{figure}
            \begin{center}
                \includegraphics[width=0.45\textwidth]{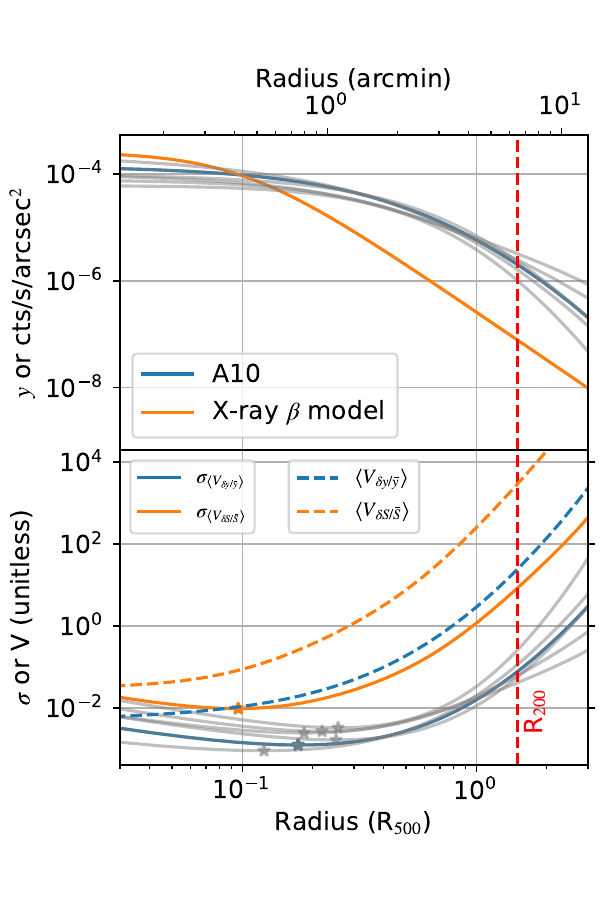}
            \end{center}
            \caption{\textbf{Top:} surface brightness profiles for an example cluster; grey lines are gNFW pressure profiles with parameter sets other than that of the A10.
            \textbf{Bottom:} the variance within a circle of radius $r$ (dashed lines) as expressed in Equations~\ref{eqn:circular_var} and \ref{eqn:average_xvar}; for the SZ profiles ($\langle V_{\delta y / \bar{y}} \rangle$ and $\sigma_{\langle V_{\delta y / \bar{y}} \rangle}$) have coloring matching that in the top panel. Each star marks the minimum of the profile upon which it is marked.}
        \label{fig:SB_Var_SpecUnc_Profiles}
        \end{figure}

%%%%%%%%%%%%%%%%%%%%%%%%%%%%%%%%%%%%%%%%%%
        At this point it is instructive to consider an example cluster with implications that can largely be generalized under the assumption of self-similarity. I take again a cluster at $z = 0.3$ with $M_{500} = 6 \times 10^{14} M_{\odot}$ and the A10 pressure profile. 
        %I fit the A10 profile with a $\beta$-model (shown in Figure~\ref{fig:SB_Var_SpecUnc_Profiles}) for an ancillary investigation which is explored in Section~\ref{sec:BetaSBs}. 
        I also explore various other pressure profiles (sets of gNFW parameters) presented in \citet{arnaud2010} and \citet{mcdonald2014}, which have corresponding profiles in grey in Figure~\ref{fig:SB_Var_SpecUnc_Profiles}.\footnote{Because I chose $z=0.3$, I adopt the parameters in the low-$z$ sample of \citet{mcdonald2014}.} 

        For the X-ray $\beta$-model:
        \begin{equation}
            I = \frac{I_0}{(1 + (r/r_c)^2)^{x}},
            \label{eqn:beta_sb}
        \end{equation}
        I adopt $\beta = 0.66$, and $r_c = 0.1R_{500}$ and normalized to be consistent with the luminosity and temperature scaling relation presented in \citet{bulbul2019}. These parameters give a qualitatively good agreement with surface brightness profiles presented in, for example, \citet{bartalucci2023,lyskova2023,romero2024}. A background rate is taken that is similar to that seen in \citet{romero2023,romero2024}.

        Adopting an \textit{XMM}-like exposure (in the EPIC cameras) and a uniform sensitivity map for SZ ($V_{y}(\theta) = \text{constant}$), the bottom panel of Figure~\ref{fig:SB_Var_SpecUnc_Profiles} show the variance profiles and associated $\sigma_{P_{\text{n,m}}}(k,\theta)$ for SZ and X-ray for arbitrary $k$ and exposure times, as neither $k$ nor exposure times affect the shape of the profiles.
    %At high $k$, we expect $P_{\rm tot}(k) \simeq P_{\rm n,m}(k)$. 

    \subsection{Incorporating biases into uncertainties}

        From Section~\ref{sec:accuracy}, we'll recall that the measured power spectra can suffer from multiple biases, which we can aggregate into a bias for each component. That is, $B_{\rm s}$ and $B_{\rm n}$ can account for the total biases on the measured signal and noise power spectra such that $P_{\rm s,m}(k) = P_{\rm s,t}(k) B_{\rm s}$ and $P_{\rm n,m}(k) = P_{\rm n,t}(k) B_{\rm s}$, for true (unbiased) spectra $P_{\rm s,t}(k)$ and $P_{\rm n,t}(k)$. Rewriting Equation~\ref{eqn:PS_signal_noise}, we have:
        \begin{equation}
            P_{\rm s,t}(k)  = \frac{P_{\rm tot,m}(k) - P_{\rm n,t}(k) B_{\rm n}}{B_{\rm s}}.
            \label{eqn:PS_true_signal}
        \end{equation}
        For $B_{\rm n} \approx 1$, one can infer a bias-corrected uncertainty on the signal power spectrum as $\sigma_{P_{\rm s,t}(k)} =  \sigma_{P_{\rm s,m}(k)} /B_{\rm s} $.

\section{Synthesizing accuracy and uncertainty concerns}
\label{sec:synthesis}

    The previous two sections have addressed matters of accuracy and uncertainty in recovered power spectra of surface brightness fluctuations. Here, I synthesize the previous two sections with respect to choice of regions. In particular, I focus on the case circular (or elliptical) annuli about the cluster center, where the goal(s) align with the determination of properties of turbulence as a function of radius at to (at least) $R_{500}$. Moreover, I want the choice of regions to accommodate SB fluctuation analysis via multiple instruments, so that analyses, at least with respect to region choices, are more homogeneous. 

    When determining properties of turbulence as a function of radius, there is a trade-off between radial resolution (how many annular regions are used) and the dynamic range that is probed with those regions, due to masking bias and sample variance. Moreover, recovering the injection scale suggests avoiding thin annuli as injection scales are likely to occur at scales where sample variance is important \citep[e.g.][]{dupourque2023,heinrich2024}. If we want to infer some radial trend, out to $R_{500}$, we want a minimum of two regions within $R_{500}$. Adopting two regions will be most accommodating to homogeneous analyses with instruments of various spatial resolutions.
    %Adopting two regions within $R_{500}$ also accommodates a sufficient dynamic range in scales recovered for various instrument resolutions.
    
    Section~\ref{sec:uncertainties} strongly motivates choosing an inner region to be a circle of radius $R_1$, such that $0.1 R_{500} < R_1 < 0.3 R_{500}$ as that range yields a minimum in uncertainties in a power spectrum of SZ fluctuations, while $R_1 \approx 0.1 R_{500}$ yields a minimum in the uncertainties of a corresponding X-ray fluctuation power spectrum. Conversely, fluctuations within $0.1 R_{500}$ may be susceptible to SB model choices and thus omitted \citep[e.g.][]{dupourque2024}, though given the nature of area-averaging, such concerns can be mitigated with a larger central region \citep[e.g.][]{romero2024}.

    While $R_1 = 0.3 R_{500}$ may be indicated (and $R_2 = R_{500}$) for the above motivations, I propose adopting $R_1 = 0.4 R_{500}$. The increase in uncertainty (relative to $R_1 = 0.3 R_{500}$), especially for SZ measurements, is not particularly consequential. In comparison, $R_1 = 0.4 R_{500}$ accommodates homogeneous analysis with a wider range of instrument resolutions and attenuates the impact of fluctuations in the central most regions. As an added bonus, $0.4 R_{500}$ is a decent approximation to $R_{2500}$ (e.g. \citet{arnaud2010} takes $R_{2500} \sim 0.44 R_{500}$). 
    %$0.4 R_{500}$ also appears to be a milepost for X-ray surface brightness profile scatter, within which profiles differ dramatically by morphologically relaxed or disturbed clusters \citep[e.g.][]{bartalucci2023}. Note that from Figure~\ref{fig:SB_Var_SpecUnc_Profiles}, the uncertainties do not increase dramatically if choosing a circular region out to $0.4 R_{500}$.

    Finally, one may consider measurements beyond $R_{500}$, for which I propose $R_{200} \sim 1.5 R_{500}$ \citep[e.g.][]{romero2020} as it satisfies the above concerns of biases and dynamic range recoverable, and of course $R_{200}$ is a useful point of comparison.

\section{Forecasting constraints on the hydrostatic mass bias: approach}
\label{sec:approach}

    In the remainder of the paper, I will turn my attention to the what constraints on the (turbulent) hydrostatic mass bias we may expect with some current and proposed future facilities. In particular, I wish to determine the redshifts and masses of clusters for which constraints of a given level can be achieved. To this end, I explore clusters between $0.05 \leq z \leq 1.65$ and $0.08 < (M_{500} / 10^{14} M_{\odot} < 12.6$.
    I adopt the annular regions suggested in Section~\ref{sec:synthesis}. Because the dominant biases when using the A12 method depend on the spectral shape, but I do not wish to assume a spectral shape nor explore a large parameter space, I calculate power spectral values assuming perfect accounting of biases. Of more concern are the associated uncertainties (detailed in Appendix~\ref{sec:HMBU}), where I do incorporate a correction to the PSF bias in the spectral uncertainties.

    \subsection{Formulation of Hydrostatic Bias}
    \label{sec:HMB}

        To relate the amplitude spectrum to the turbulent Mach number, $\mathcal{M}$, I use the (adapted) relation from \citet{Gaspari2013_PS}:
        \begin{equation}
            \mathcal{M} = 4 A_{\rho} \left( \frac{l_{\rm inj}}{0.4 R_{500} } \right )^{-\alpha_H} = 2.4 A_{P} \left( \frac{l_{\rm inj}}{0.4 R_{500}} \right)^{-\alpha_H},
            \label{eqn:mach_relation}
        \end{equation}
        where $A_{\rho}$ and $A_{P}$ are the peaks of the density and pressure amplitude spectra, respectively. The injection scale is denoted as $l_{\rm inj}$ and $\alpha_H \simeq 0.25$ is an empirically derived scaling for the hydrodynamical cascade of fluctuations. This scaling relation has been found to be in agreement with with the relation from \citet{zhuravleva2012}; see, e.g. \citet{romero2023,dupourque2024}.

        To determine the input turbulent Mach numbers I employ its relation to the non-thermal pressure fraction due to turbulent motions:
        \begin{equation}
            \frac{P_{\rm NT}}{P_{\rm th}} = (\gamma/3) \mathcal{M}^2,
            \label{eqn:PNT-Mach}
        \end{equation}
        where I take $\gamma = 5/3$ to be the adiabatic index. With this relation, I use the radial profile of non-thermal pressure as given in \citet{nelson2014a}:
        \begin{equation}
            \frac{P_{\rm NT}}{P_{\rm th} + P_{\rm NT}} = 1 - A \left( 1 + exp\left[ - \left(\frac{r/r_{200m}}{B}\right)^{\gamma_P} \right] \right),
        \end{equation}
        where they found values of $A = 0.452 \pm 0.001$, $B=0.841 \pm 0.008$, and $\gamma_P = 1.628 \pm 0.019$. 
        While \citet{nelson2014a} and \citet{battaglia2012a} consider the non-thermal pressure to be completely from random (i.e. turbulent) motions, other sources of non-thermal pressure may not be negligible \citep[e.g.][]{angelinelli2020}.
        
        %An extension of constraining turbulent velocities is to derive the non-thermal pressure support from said velocities and the consequent hydrostatic mass bias (which may not be the total hydrostatic mass bias \citep[e.g.][]{angelinelli2020}).
    
        \begin{figure}[!h]
            \begin{center}
                \includegraphics[width=0.45\textwidth]{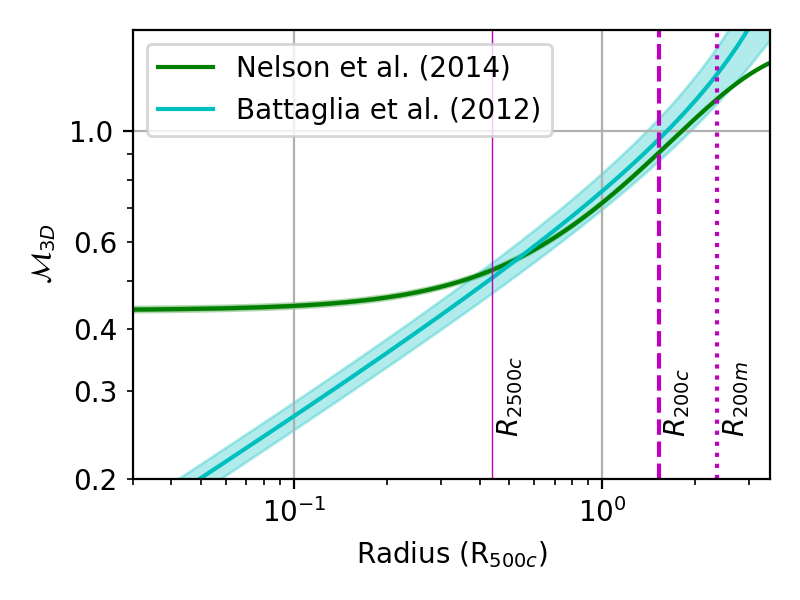}
            \end{center}
            \caption{Mach profiles, as inferred from Equation~\ref{eqn:PNT-Mach} and non-thermal pressure profiles in \citet{battaglia2012a} and \citet{nelson2014a}.}
            \label{fig:MachProfiles}
        \end{figure}

        \citet{khatri2016} provide a relation between the hydrostatic bias and $\mathcal{M}_{\text{3D}}$ (and attach a corresponding subscript to denote the method of calculation, $b_{\mathcal{M}}$): %\MG{there should be a minus in front, added}:
        \begin{equation}
            b_{\mathcal{M}} = \frac{-\gamma \mathcal{M}_{\text{3D}}^2}{3} \frac{ d \ln P_{\text{NT}}}{d \ln P_{\text{th}}} \left( 1 +  \frac{\gamma \mathcal{M}_{\text{3D}}^2}{3}\frac{ d \ln P_{\text{NT}}}{d \ln P_{\text{th}}} \right)^{-1},
            \label{eqn:mach_bias}
        \end{equation}
        where $\gamma$ is the adiabatic index, taken to be 5/3 for the ICM, and $b_{\mathcal{M}}$ is implicitly a function of $r$. NB that as defined in \citet{khatri2016} $b_{\mathcal{M}} \equiv M_x / M_{\text{tot}} - 1$. Following the recasting performed in \citet{khatri2016}, I find:
        \begin{equation}
            \frac{ d \ln P_{\text{NT}}}{d \ln P_{\text{th}}} = \frac{ d \ln P_{\text{NT}} / d \ln r}{d \ln P_{\text{th}} / d \ln r} = 1 + 2 \frac{ d \ln \mathcal{M}_{\text{3D}} / d \ln r}{d \ln P_{\text{th}} / d \ln r}.
            \label{eqn:NTfrac}
        \end{equation}

        I adopt the rings suggested in Section~\ref{sec:synthesis}; those with edges at [0.4, 1.0, 1.5] $R_{500}$. The effective radii for these rings is roughly [0.3, 0.8, 1.3] $R_{500}$. To calculate $b_{\mathcal{M}}(R_{2500})$, $b_{\mathcal{M}}(R_{500})$, and $b_{\mathcal{M}}(R_{200})$, I need to estimate $\mathcal{M}(R_{2500})$, $\mathcal{M}(R_{500})$, and $\mathcal{M}(R_{200})$. From the inner two rings, or all three rings, I fit a power law slope to $\mathcal{M}(r)$ at the effective radii of each ring and use the fitted power law to calculate $\mathcal{M}(R_{2500})$, $\mathcal{M}(R_{500})$, and $\mathcal{M}(R_{200})$ (and consequently $b_{\mathcal{M}}(R_{2500})$, $b_{\mathcal{M}}(R_{500})$, and $b_{\mathcal{M}}(R_{200})$), propagating all uncertainties, including those from sample variance (see Appendix~\ref{sec:HMBU}).

        %Nonetheless, the objective is then to calculate $b_{\mathcal{M}}$ at radii of interest, where I consider the primary radii of interest to be $R_{2500}$, $R_{500}$, and $R_{200}$. Therefore, $\mathcal{M}_{\text{3D}}(r)$, $d \ln \mathcal{M}_{\text{3D}} / d \ln r$, and $d \ln P_{\text{th}} / d \ln r$ should each be calculated at $R_{2500}$, $R_{500}$, and $R_{200}$. I have already established that a single power law will be assumed for $\mathcal{M}_{\text{3D}}(r)$, and thus one can use the fitted power law to calculate $\mathcal{M}_{\text{3D}}$ at the desired radii. 
        
        Equation~\ref{eqn:NTfrac} also requires calculation of $d \ln P_{\text{th}} / d \ln r$, for which I assume that the logarithmic slope from the A10 profile is recovered. 
        Relative to the uncertainties on other quantities in Equation~\ref{eqn:NTfrac}, this should not be the limiting factor, even for X-ray observation. Given that the uncertainty for this quantity in the case SZ observations is straightforward (see below), I include uncertainties in the SZ case. In the X-ray case, this uncertainty is ignored.

    \subsection{Toy setup}

        In order to quantify the expected uncertainties on turbulent hydrostatic mass bias, I require a SB model, a model of the noise (spatially and spectrally), and (3) a defined region and scale being sampled for a given observation (Section~\ref{sec:uncertainties}). 
        %In short, I am able to calculate inferred uncertainties based on the above information without directly requiring the generation of noise realizations, and in some cases without mock observations.

        \subsubsection{SZ modelling}

            To forecast all SZ observations, I again adopt the A10 pressure profile and I assume white noise, often with a radial dependence. Mapping speeds can be calculated from the various time sensitivity calculators provided for 
            MUSTANG-2\footnote{\url{https://greenbankobservatory.org/science/gbt-observers/mustang-2/}},  
            NIKA2\footnote{\url{https://publicwiki.iram.es/Continuum/TimeEstimatorScriptGuideW2023}}, 
            TolTEC\footnote{\url{https://toltec.lmtgtm.org/toltec_sensitivity_calculator}}, 
            and AtLAST\footnote{\url{https://github.com/ukatc/AtLAST_sensitivity_calculator}}.
            WIKID is a proposed successor to MUSTANG-2; I have produced mapping speed estimates using MUSTANG-2 data with the projected increased detectors and improved detector and readout noise. I assume that the instruments perform scans which are roughly equivalent to Lissajous daisy scans of $5{\arcmin}$ (NIKA2), $3{\arcmin}$ (TolTEC), $3{\arcmin}.5$ (MUSTANG-2), and $3{\arcmin}.5$ (WIKID), on a single pointing. 
            %These scan sizes are plausibly well-performing scan sizes given each instrument's instantaneous field of view. The corresponding mapping speeds in Compton $y$ are calculated assuming a direct conversion from the mapping speeds in units of mJy/beam in either the 150 GHz band (for NIKA2 and TolTEC) or the 90 GHZ (for MUSTANG-2 and WIKID).

            AtLAST\footnote{\url{https://www.atlast.uio.no/}} is intended to have a camera with over a degree field of view; currently the aim is 2 degrees. 
            %While I might still consider adopting a transfer function and radial noise pattern, I consider that for our scales AtLAST has a constant transfer function of unity and a uniform noise map. The achieved noise is calculated via their sensitivity calculator assuming ALMA-like bands. 
            Assuming an optimal internal linear combination (ILC) of sensitivites in ALMA-like bands, I find a mapping speed of $1.6\times 10^{-6}$-root hour in Compton y, similar to that found in \citet{dimascolo2024}. For comparison, SPT-SZ \citep{bleem2015} would achieve an optimal ILC sensitivity of $\sim3 \times 10^{-6}$ in Compton $y$-arcmin, but their minimum variances actually achieve $1.4\times10^{-6}$ in Compton $y$-arcmin. This corresponds to a degradation factor of $4.8$. While the sensitivities of each band and angular resolution will affect this degradation, I adopt this factor and arrive at a Compton $y$ mapping speed of $8 \times 10^{-6}$-root hour at beam scales. 

            SPT-3G is expected to be $\sim$ten times more sensitive than SPT-SZ \citep{benson2014}. For both SPT-3G and AtLAST will employ a survey strategy such that the noise is uniform and the transfer function is effectively unity at scales being probed. I utilize a parametric model of the transfer function fit to the MUSTANG-2 transfer function \citep{romero2020}, and I scale this model to the FOV of NIKA2, TolTEC, and WIKID. 

        \subsubsection{X-ray modelling}

            For the case of the X-ray observations, I assume an isothermal $\beta$ model (Equation~\ref{eqn:beta_sb}) with $\beta = 2/3$, $r_c = 0.1 R_{500}$ and $I_0$ taken to satisfy the $L_{X,cin}{–}M500{–}z$ and $T_{X,cin}{–}M500{–}z$ relations in \citet{bulbul2019}, which is seen to be qualitatively similar to to profiles in the literature (as mentioned in Section~\ref{sec:uncertainties}).
            To properly normalize the SB profiles, I employ \lstinline{SOXS}\footnote{\url{https://hea-www.cfa.harvard.edu/soxs/index.html}}\citep{soxs2023} which calculates the rest-frame luminosity based on temperature, and from this I derive the appropriate SB normalization. 
            
            I then use \lstinline{SOXS} to simulate observations, with a single pointing, in the 0.5 to 2.0 keV range of our X-ray (imaging) instruments on \textit{Lynx}, \textit{NewAthena}, \textit{AXIS}, and \textit{LEM}. I assume a modest column density (4e18 cm$^-2$) on all observations. I apply instrumental background and foreground to one observation per instrument to calculate a background count rate. I use \lstinline{pyproffit} to extract SB profiles, and fix $\beta = 2/3$ and $r_c = 0.1 R_{500}$.
            %I recover the remaining profile parameters. I take the background rate fit to the one observation (per instrument) and apply this background rate to all other observations of that instrument.
            
            For \textit{XMM}, I use results obtained in \citet{romero2024} to establish a scaling relation with self-similar scaling: $L_X \propto M_{500} E(z)^2$ that corresponds to the soft energy band (0.5 to 1.25 keV) used in \citet{romero2024}. As the background, I take an average of the total backgrounds in the 0.5 to 1.25 keV and assume this background is constant for our forecasting constraints. This creates a slight mismatch, but I do not intend to focus on the forecast results for \textit{XMM}; I include them to show consistency, in a very broad sense, between my forecasts and results which have been published. 
            
        %\subsubsection{Regions sampled and uncertainty calculation}
%
%            As proposed in Section~\ref{sec:spectral_uncertainties}, I opt to use rings with radial bounds at 0, 0.4, 1.0, and 1.5 $R_{500}$. For a given scale, our inferred power spectrum uncertainty is given by Equation~\ref{eqn:sample_variance}, where $P_n(k)$ is the power spectrum of our noise. In particular, this will be given by our assumed white noise divided by the SB model. If I am investigating a targeted SZ observation, then I also account for the transfer function in $P_n(k)$. Our total 2D power spectrum uncertainty is given by:
%            \begin{equation}
%                \sigma_{P_{\rm 2D}} = ( \sigma_{P_n}^2 + \sigma_{\rm SV}^2 )^{1/2}.
%            \end{equation}
%            
%            $N_{\rm eff}$ are calculated as in Section~\ref{sec:accuracy} for each ring and allow me to deproject to $P_{\rm 3D}$ and $\sigma_{P_{\rm 3D}}$. To calculate $\sigma_{A_{\rm 3D}}$, I %have:
%            \begin{equation}
%                \sigma_{A_{\rm 3D}} = \frac{\sigma_{P_{\rm 3D}}}{2 P_{\rm 3D}} A_{\rm 3D}.
%            \end{equation}
%            As specified earlier in this section, I have values for $P_{\rm 3D}$, and consequentially $A_{\rm 3D}$, from our assumptions about the non-thermal pressure profile and scaling relation between Mach number and amplitude at the peak of the amplitude spectrum. To reiterate, when I sample different scales, I am always assuming that the scale in question is the injection scale. 
            %{eqn:mach_relation}

        \subsubsection{Remaining parameters}
        \label{sec:varied_params}

            In order to employ Equation~\ref{eqn:mach_relation}, I must adopt an injection scale. However, as the literature has not borne out a clear expectation for what the injection scales will be across a large range of redshifts and masses. Observational and numerical simulations tends to find injection scales of a few hundreds of kpc \citep[in the masses, redshifts probed; e.g.][]{gaspari13,khatri2016,sanders2020,romero2023,dupourque2023,dupourque2024}. Additionally, the injection scale is expected increase with cluster radius. Therefore, I consider a range of injections scales:  $[0.005,0.008,0.013,0.025,0.05,0.1,0.2,0.4,0.6,0.8] \times [R_{500},1 $ Mpc], where injection scales $<0.1 \times [R_{500},1 $ Mpc] are included for the purposes of cataloging the uncertainties in the recovered spectra at those scales (see Section~\ref{sec:discussion}). I consider plausible injection scales to be $\geq 0.1 \times R_{500}$.

            Finally, to account for the noise in maps with targeted instruments, I must assume an exposure time. Although AtLAST is likely to perform a survey \citep{dimascolo2024}, the depth of that survey is yet to be determined. As such, I explore various exposure times for it as well. In particular, I consider three exposure times: 50, 250, and 1000 ks. While 50 ks is still not a short exposure, in the regime of SB fluctuations, it should be considered a short exposure. 250 and 1000 ks thus correspond to a modest and deep exposure. 

            %%% Discuss what has been seen in the literature...

\section{Forecasting constraints on the hydrostatic mass bias: results}
\label{sec:results}

    \begin{figure}[!h]
        \begin{center}
            \includegraphics[width=0.45\textwidth]{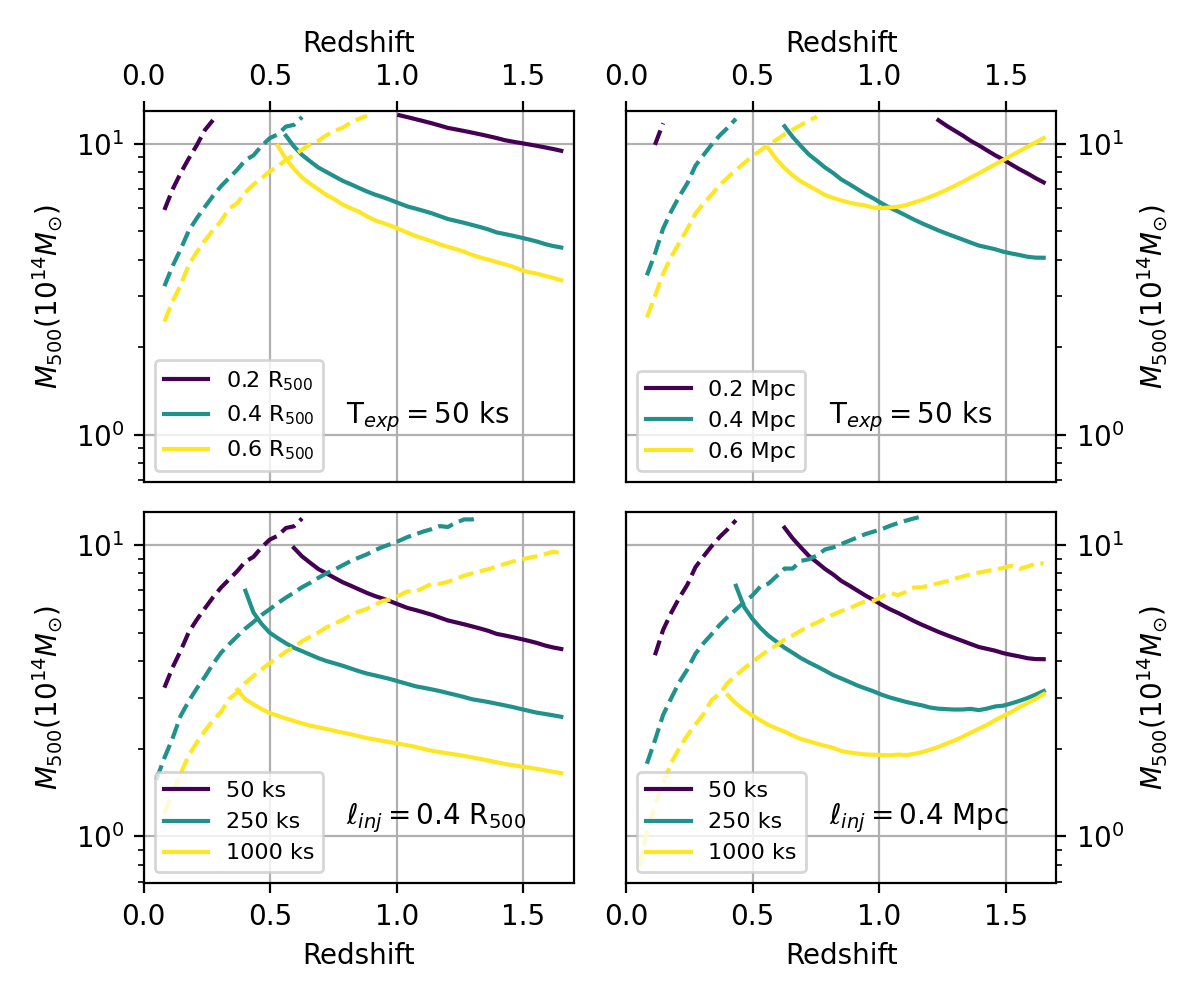}
        \end{center}
        \caption{Mass thresholds for achieving $\sigma_{b_{\mathcal{M}}} = 0.1$ for MUSTANG-2 (solid lines) and \textit{XMM} (dashed lines). Above these lines the uncertainties on $b_{\mathcal{M}}$ will be less than $0.1$. The various panels in this figure show how the constraints change based on assumptions and exposure time. \textbf{Top left:} The injection scale is assumed to scale with $R_{500}$ and all results take a fixed exposure time (on-source) of 50 ks. \textbf{Top right:} same as top left, but injection scales are at specified physical scales. \textbf{Bottom left:} Injection scale is held at a fixed fraction of $R_{500}$ and exposure times are varied. \textbf{Bottom right:} Same as bottom left except the injection scale is held at a fixed physical scale.}
        \label{fig:constraints_vs_pars}
    \end{figure}

    As discussed in the previous section, SB profiles have been assumed for clusters between $0.05 \leq z \leq 1.65$ and $0.08 < (M_{500} / 10^{14} M_{\odot} < 12.6$. Accounting for expected image noise, imaging fields of view (for a single pointing), transfer functions, and a Gaussian PSF assumption, I can calculate the mass threshold (as a function of redshift) above which instruments can achieve a given uncertainty in the hydrostatic mass bias, $\sigma_{b_{\mathcal{M}}}$ or better. Given that values for the hydrostatic mass bias, $b$, are expected to be $0.1 \lesssim b \lesssim 0.3$\citep[e.g.][]{hurier2018}, a comparable (or better) can be taken as a target uncertainty. I take $\sigma_{b_{\mathcal{M}}} = 0.1$ as my target uncertainty (on a single cluster) as this appears achievable in the presented approach, while uncertainties much less than this may not be possible due to sample variance.
    
    \begin{figure}[!h]
        \begin{center}
            \includegraphics[width=0.48\textwidth]{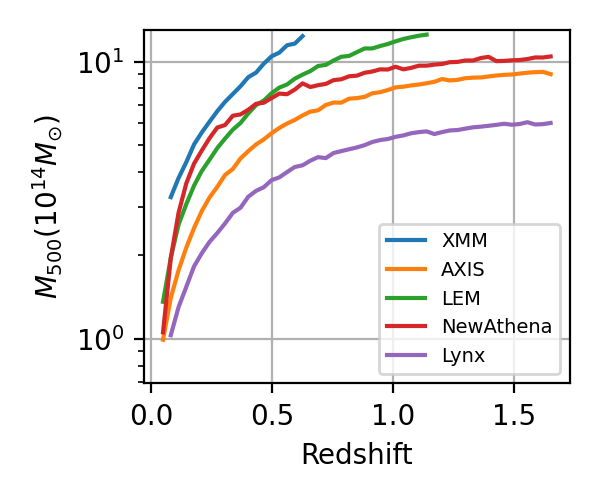}
            \includegraphics[width=0.48\textwidth]{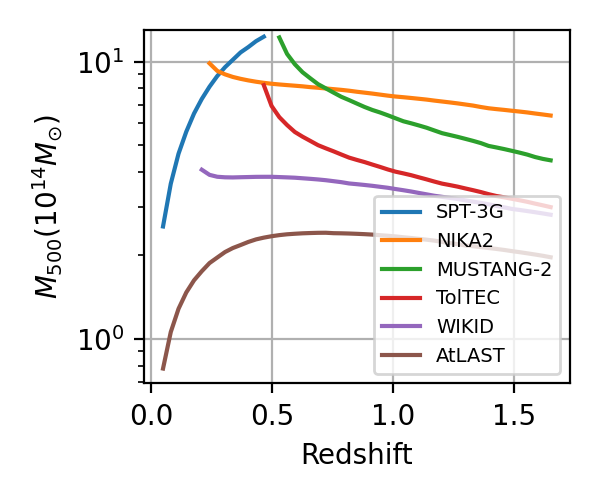}
        \end{center}
        \caption{Mass thresholds (as in Figure~\ref{fig:constraints_vs_pars} for achieving $\sigma_{b_{\mathcal{M}}} = 0.1$ when $l_{\rm inj} = 0.4 R_{500}$ for select current and proposed X-ray facilities (\textbf{top}) and SZ facilities (\textbf{bottom} with 50 ks on-source time. Here, I assume imaging cameras are used. Limitations at the low-redshift end come from $R_{500}$ exceeding the field of view. The inferred SPT-3G results assume the end-of-survey expected sensitivity.}
        \label{fig:HyMaBias_Xray_SZ}
    \end{figure}
    
    %Even though PSF corrections are applied (assuming $\alpha=3$), most results presented in this section correspond to angular scale larger than most PSFs of the instruments explored. I do not incorporate other potential corrections noted in Section~\ref{sec:accuracy} as those present at larger scales do not significantly change the inferred amplitude and their corrections would apply knowledge of an underlying spectrum. Rather than produce results valid for a specific spectrum, I focus on calculating the uncertainty purely due to noise in the image and stochasticity of fluctuations. In this manner, our calculations should serve as a floor (best-case scenario) for estimating what constraints may be achieved when estimating the hydrostatic mass bias of clusters at $R_{500}$. 

    Figure~\ref{fig:constraints_vs_pars} shows how the mass threshold to achieve $\sigma_{b_{\mathcal{M}}} = 0.1$ changes as a function of $l_{\text{inj}}$ and exposure (integration) time, $T_{\rm exp}$. The improvement in constraints as the injection scale increases can be inferred as a consequence of how the noise varies with scale (Secton~\ref{sec:uncertainties}) and how quickly the signal varies with scale (e.g. $\alpha \gtrsim 3$).
    
    \begin{figure}[!h]
        \begin{center}
            \includegraphics[width=0.48\textwidth]{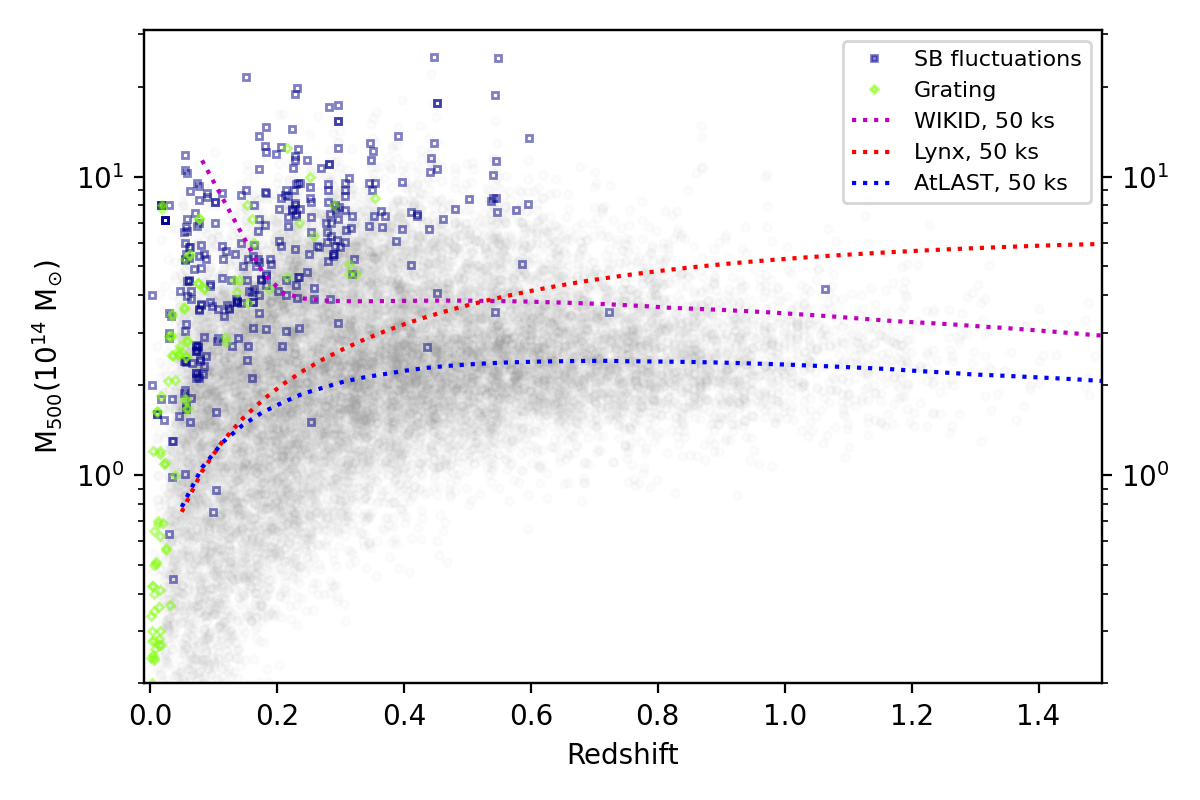}
        \end{center}
        \caption{Mass thresholds for achieving $\sigma_{b_{\mathcal{M}}} = 0.1$ when $l_{\rm inj} = 0.4 R_{500}$ for some of the more sensitive X-ray or SZ facilities. The grey dots are clusters in ACT, SPT, Planck, MCXC, or eRosita catalogs and serve to contextualize the mass threshold curves. Moreover, the blue boxes and chartreuse diamonds indicate masses and redshifts of clusters (or elliptical galaxies) studied via SB fluctuations or spectroscopy to infer gas velocities, but generally not out to $R_{500}$.} 
        \label{fig:HyMaBias_BestOf}
    \end{figure}
    
    While I explored the consequence of assuming injections scales at fixed physical scales, I anticipate that injection scales that scale with $R_{500}$ are likely to be the better descriptor of fluctuations in clusters. As such, Figure~\ref{fig:HyMaBias_Xray_SZ} adopts the case that $l_{\text{inj}} = 0.4 R_{500}$ and $T_{\text{exp}} = 50$ ks, with the exception of SPT-3G, whose sensitivity (noise) is the expected noise at the end of the survey.

    Not surprisingly, one of the more striking results is that SZ instruments can perform much better at higher redshift. Conversely, several SZ instruments are quite hampered at low redshift due to their instantaneous field of view, which hurts both the imageable field of view and most consequentially, the transfer function. While SZ survey instruments have sufficiently large fields of view, their resolution and sensitivity limit the constraints they can provide. I note that results for SO and CMB-S4 are similar as SPT-3G as the expected sensitivities are similar with SO and CMB-S4 having marginally poorer angular resolution than SPT-3G.

    Figure~\ref{fig:HyMaBias_BestOf} retains forecasts of $\sigma_{b_{\mathcal{M}}}$ for just three instruments. For context, I have added clusters detected in several catalogs. For additional context, I have added clusters, groups, and/or BCGs for which turbulent velocities have been measured or inferred. Most of studies of turbulent velocities, either by spectroscopy or SB fluctuations, have focused on the central regions of clusters, and several clusters have been studied multiple times. That is to say that the constraints denoted by dotted curves in Figure~\ref{fig:HyMaBias_BestOf} signify a substantial improvement in the existing constraints of fluctuations out to $R_{500}$.

    %\textcolor{red}{I need to add in a mention of angular resolution here...}

\section{Discussion}
\label{sec:discussion}

\begin{figure*}
        \begin{center}
            \includegraphics[width=0.98\textwidth]{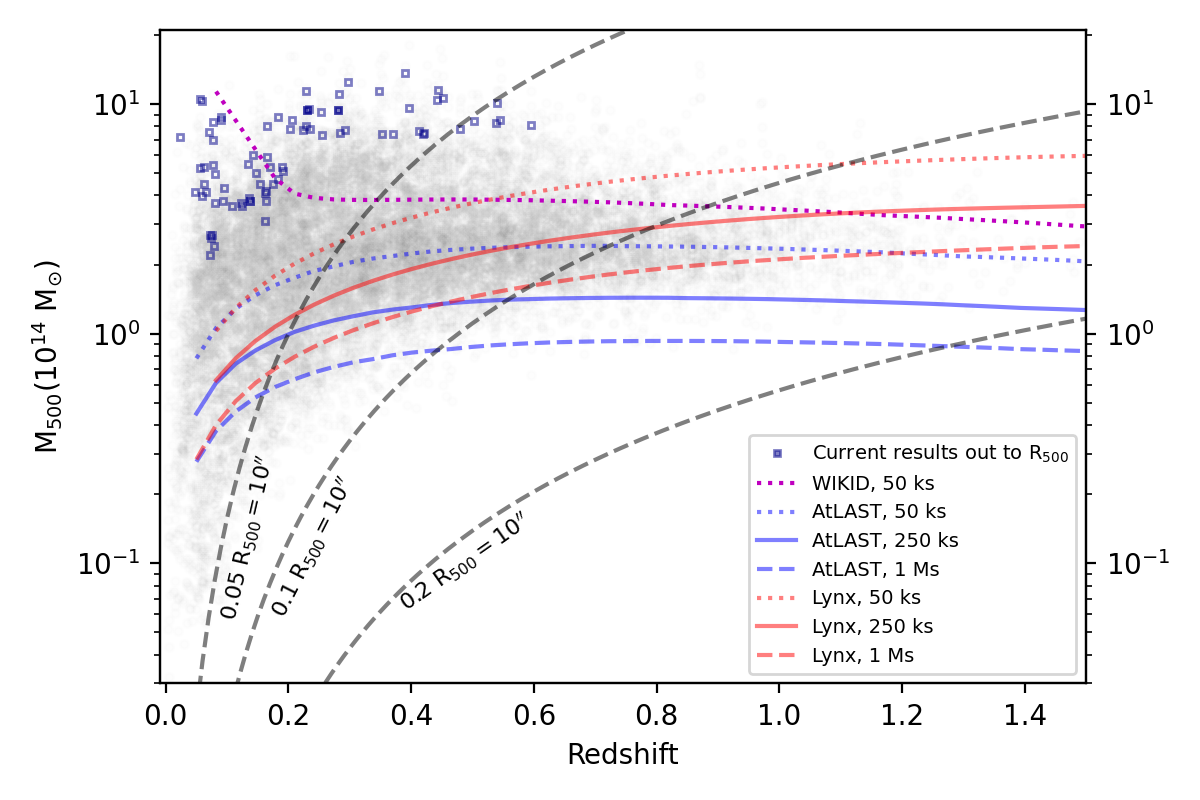}
        \end{center}
        \caption{Similar to Figure~\ref{fig:HyMaBias_BestOf} with thresholds for achieving $\sigma_{b_{\mathcal{M}}} = 0.1$ on $M_{500}$ when $l_{\rm inj} = 0.4 R_{500}$ by instrument and exposure time. Here I have limited markers to those which indicate some (often poor) constraints on turbulent velocities out to $R_{500}$ from SB fluctuation analyses, some of which have corresponding hydrostatic mass bias estimates.} 
        \label{fig:HyMaBias_R500}
    \end{figure*}

\begin{figure*}
        \begin{center}
            \includegraphics[width=0.98\textwidth]{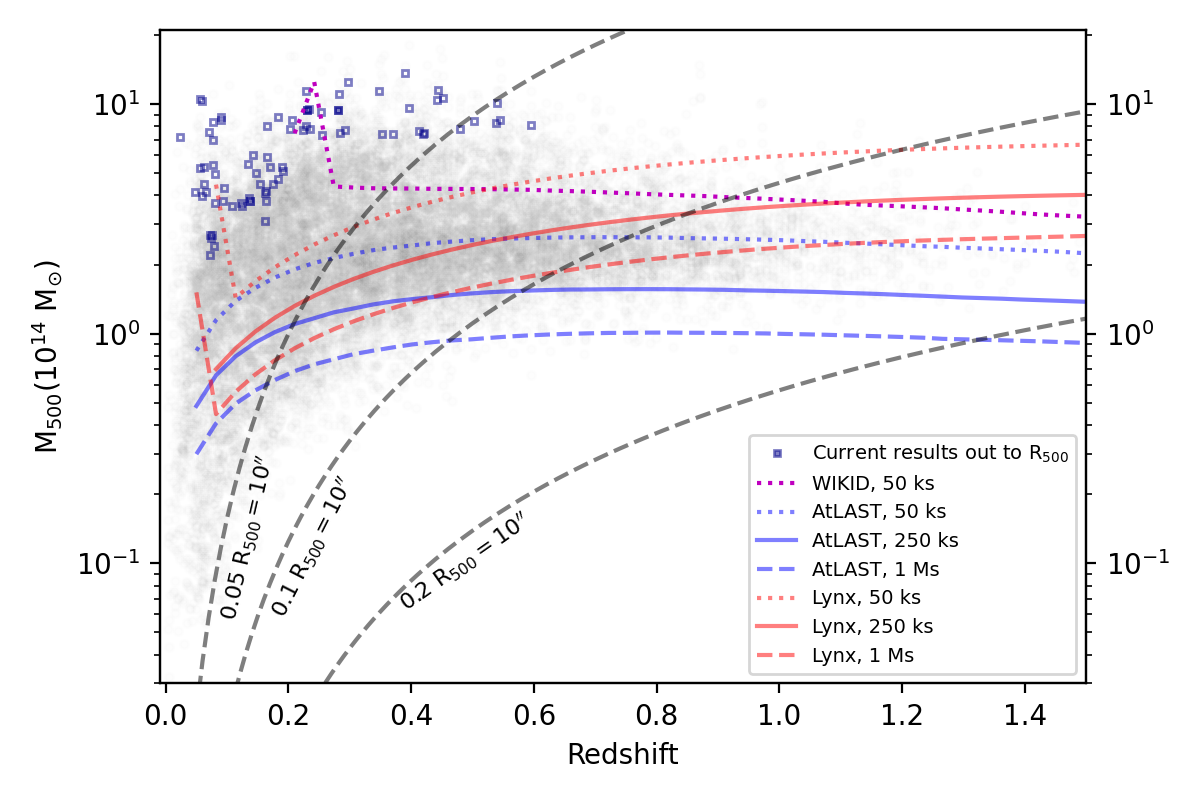}
        \end{center}
        \caption{Similar to Figure~\ref{fig:HyMaBias_R500}, but curves represent thresholds for achieving $\sigma_{b_{\mathcal{M}}} = 0.1$ on $M_{200}$.} 
        \label{fig:HyMaBias_R200}
    \end{figure*}

    Section~\ref{sec:results} presented constraints on the hydrostatic mass bias assuming perfect bias corrections and that a single measurement at the injection peak was sufficient to infer the turbulent velocities and subsequently the hydrostatic mass bias. While decent bias corrections can be achieved, doing so requires access to nearly a decade in scales (Section~\ref{sec:plausible}). This requirement, along side the secondary smoothing bias (Section~\ref{sec:smooth_remarks}) suggest the need for the FWHM of an instrument to be less than $0.1 R_{500}$ for a targeted cluster. Figure~\ref{fig:HyMaBias_R500} adds curves denoting what fraction of $R_{500}$ corresponds to $10^{\prime\prime}$. Additionally, constraints for the three integration times are shown for AtLAST and \textit{Lynx}. Finally, in Figure~\ref{fig:HyMaBias_R500}, I have trimmed the studies in the literature to those with constraints (even if marginal) out to $R_{500}$. Figure~\ref{fig:HyMaBias_R200} replicates Figure~\ref{fig:HyMaBias_R500}, but shows constraints for $\sigma_{b_{\mathcal{M}}} = 0.1$ at $R_{200}$. 

    Two prominent findings from Section~\ref{sec:plausible} were that (1) assuming the injection scale is within the scales probed, good constraints on the bias-corrected spectral shape benefit greatly when $k_{\rm peak, raw} \lesssim k_{\rm max}/4$ and (2) the peaks of the raw amplitude spectra, $A_{\rm 3D}(k_{\rm peak})$, and the overall density or pressure fluctuations, $\sigma_{\rm 3D}$, as recovered in the raw spectra are not substantially biased. 
    %Notwithstanding bias corrections, recovering scales $\gtrsim 4$ times smaller than $l_{\rm inj}$ should recover $\gtrsim90$\% of the integrated spectrum, assuming $\alpha = 11/3$. To ensure a high level of confidence in the inferred turbulent Mach numbers via either relation (from the peak or integrated spectrum) it is therefore advisable to seek at least a factor of 4 better resolution (e.g. FWHM) relative to the injection scale. 
    In the case that $l_{\rm inj} = 0.4 R_{500}$, then the goal is to sufficiently recover fluctuations at scales of $0.1 R_{500}$. In this respect, resolution should never be a limiting factor \textit{Lynx}. However, for instruments such as WIKID or AtLAST, with FWHM of $10^{\prime\prime}$, then resolution is a concern, where the grey line marked $0.1 R_{500} = 10^{\prime\prime}$ suggests a a threshold beyond which AtLAST or WIKID may not provide robust constraints.
    
    %Focusing on the case of AtLAST: if the injection scale happens to be larger than $0.4 R_{500}$, this could potentially ameliorate the recovery of the power spectrum shape. However, in Ring 1, scales larger than $0.4 R_{500}$ will begin to suffer heavily from sample variance.

    On account of the injection scale, the requirement to obtain fluctuations down to $0.1 R_{500}$ may be alleviated if the injection scale is larger than $0.4 R_{500}$, which is quite plausible for Rings 2 and 3. However, there remains the challenge of recovering down to $0.1 R_{500}$ in Ring 1 (assuming $l_{\rm inj} \simeq 0.4 R_{500}$). If the secondary bias due to beam smoothing (Section~\ref{sec:smooth_remarks}) can be satisfactorily addressed, then the principal remaining concern is what sensitivity is required at $4 k_{\rm peak, raw}$ to satisfactorily recover the unbiased power spectrum. Appendix~\ref{sec:ShapeRecovery} suggest that $\sigma_{A_{\text{3D}}} \simeq 0.05$ is sufficient.

    \begin{figure}
        \begin{center}
            \includegraphics[width=0.45\textwidth]{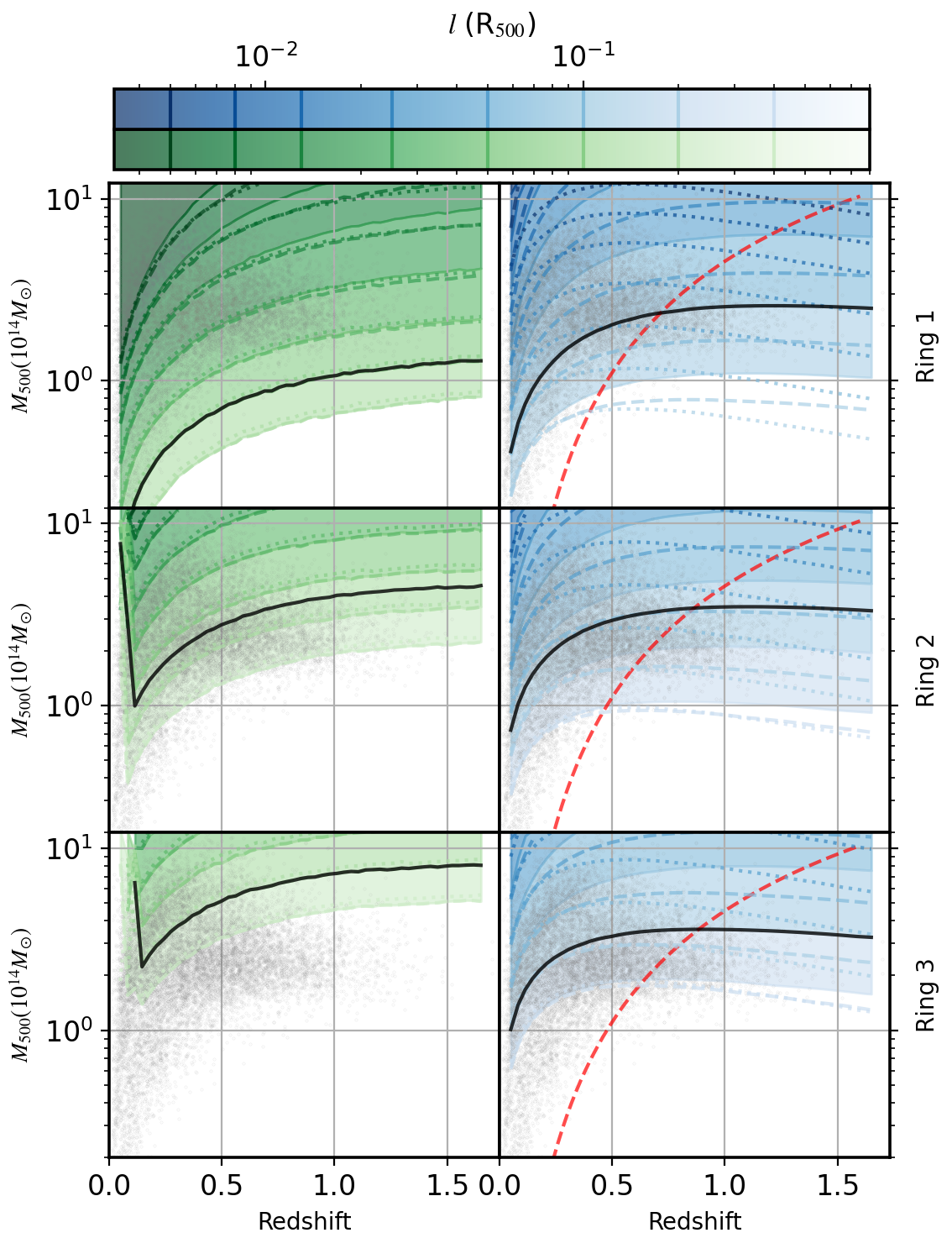}
        \end{center}
        \caption{Scales relative to $R_{500}$ for which either \textit{Lynx} (\textbf{left}, green) or AtLAST (\textbf{right}, blue) achieves an uncertainty in $A_{3D}$ of 0.05 or smaller with a 1 Ms exposure. The top row is taken for Ring 1 (0-0.4$R_{500}$); the middle row for Ring 2 (0.4-1.0 $R_{500}$); and the bottom row for Ring 3 (1.0-1.5$R_{500}$). Maximum scale at which the shading starts is $0.4 R_{500}$. Dotted curves correspond to thresholds if no PSF corrections are needed (i.e. the PSF is infinitesimal); dashed curves correspond to thresholds if the FWHM of the PSF were half of the nominal FWHM.  Black curves denote constraints at scales which could plausibly be at $l_{\rm inj}/4$ (see text). Grey dots are detected clusters from surveys, as in Figure~\ref{fig:HyMaBias_BestOf}. The red dashed curve denotes $0.1 R_{500} = 10^{\prime\prime}$.} 
        \label{fig:MZ_sigA2d-thresh}
    \end{figure}

    %\textcolor{red}{Remember to check what is required in terms of constraints at $l_{\rm inj} / 4$.}
    %If we assert that the limiting factor in accurately determination of $b_{\mathcal{M}}$ stems from constraints on $A_{3D}(4 k_{\rm inj})$, then one might ask what is a sufficient constraint on $A_{3D}(4 k_{\rm inj})$?
    %In the case that no statistical noise exists, and one only needs to worry about the sample variance, then the general shape of the spectra are well recovered with very gentle priors (e.g. priors with some pre-MCMC fits, and priors in the form of an assumed mask bias). However, if one adds noise into the mix, then I find that some priors are required. 
    %Appendix~\ref{sec:ShapeRecovery} builds on the plausible scenario explored in Section~\ref{sec:plausible} and explores the interplay between priors and noise level required to achieve a good agreement with expected spectra.  Appreciable uncertainties $\sigma_{A_{\rm 3D}}(4 k_{\rm inj}) = 0.05$ to $0.1$ still yield decent recovery of the spectral shapes. 
    
    Figure~\ref{fig:MZ_sigA2d-thresh} shows the mass thresholds where $\sigma_{A_{\rm 3D}}(k=1/l) = 0.05$ for various scales. The black curves indicate the thresholds for the (modestly conservative) case that $l_{\rm inj,2} \gtrsim 0.6 R_{500}$ in Ring 2 and $l_{\rm inj,3} \gtrsim 0.8 R_{500}$ in Ring 3.\footnote{Recall from Section~\ref{sec:approach}, the effective radii for the rings are roughly [0.3, 0.8, 1.3] $R_{500}$.} That threshold is fairly consistent across the three rings (at $M_{500} \approx 3 \times 10^{14} \text{M}_{\odot}$) for AtLAST, while the threshold shows a marked evolution across the three rings for \textit{Lynx}. 

    Figure~\ref{fig:MZ_sigA2d-thresh} also shows the constraints if the noise in the images were the same and the FWHM of the PSF were half of its nominal value (dashed curves) or were infinitesimal (dotted curves). For Lynx, the improvement (on the scales shown) is minimal, while for AtLAST the improvements are not trivial. Even so, the nature of a cascading power spectrum simply makes measurements at very small scales difficult even with an infinitesimal PSF. 
    
    %I've alluded to the expectation for $l_{rm inj}$ to increase with cluster radius, I have thus far not address the complementary potentials of observing a cluster with multiple instruments, especially instruments with different resolutions. Most prominently, we may think of an X-ray and SZ observation. Figure~\ref{fig:MZ_sigA2d-thresh} \textcolor{red}{Blah blah}.

    \subsection{Caveats}

        In the course of making these forecasts many assumptions were made, all of which will have some impact on the results. One of the most fundamental choices was to adopt the A12 method, which has two dominant biases. Even so, its flexibility seems likely to keep it as a preferred method. 

        The choice of Mach relation (that from \citet{gaspari13}) was made out of simplicity. Adopting a different relation \citep[e.g. those from][]{zhuravleva2012,zhuravleva2023} are unlikely to dramatically change the results. That is, both relations depend on decent recovery of spectra over a similar range of scales to ensure accuracy. Moreover, these two relations are found to be in good agreement \citep[e.g.][]{romero2023,dupourque2024} with respect to inferred Mach numbers from observed spectra.

        %A few observational works exist constraining the non-thermal pressure profiles \citep[e.g][]{eckert2019} and are not sufficient to determine a "universal" model. Thus, 
        To model the turbulent non-thermal pressure profile, one looks to simulations \citep[e.g][]{nelson2014a,shi2015,biffi16,angelinelli2020}, where parameterizations were provided in \citet{battaglia2012a,nelson2014a}. \citet{battaglia2012a} and \citet{nelson2014a} are in decent agreement (see Figure~\ref{fig:MachProfiles}), while \citet{angelinelli2020} finds a shallower slope similar to \citet{eckert2019}. While I took the profile from \citet{nelson2014a} as my fiducial turbulent non-thermal pressure profile, I find similar results if I use the profiles in \citet{battaglia2012a} or \citet{angelinelli2020}.

        Another assumption falls into the assumed SB profiles (or equally, pressure profiles, for SZ observations). Figure~\ref{fig:SB_Var_SpecUnc_Profiles} showed that the A10 pressure profile is similar to other pressure profiles; conversely, all sample-averaged profiles compared are quite similar. The adopted X-ray SB profile was also found to be consistent with those in \citet{bartalucci2023,lyskova2023}.

        \subsubsection{Additional Noise Concerns}
        \label{sec:noise_and_contamination}

            Another critical assumption corresponds to the noise characteristics assumed. Both SZ and X-ray are susceptible to contamination from (1) compact sources, (2) intracluster contaminants, (3) extended foreground, and (4) extended background contaminants. In the X-ray, compact sources are generally masked, while in the SZ they may be subtracted (at the cost of additional uncertainty) or masked. In the literature, the treatment of intracluster contaminants is currently ad hoc, especially as there is not an unambiguous definition of what constitutes an intracluster contaminant \citep[e.g.][]{Zhuravleva2015,romero2023}. Notwithstanding a lack of clear definition, the approach is generally to mask these contaminants.
            %(e.g. infalling group or other identifiable structure that does not appear to be generated from turbulence).
            
            It may be possible that some extended foregrounds or backgrounds can be fit out (e.g. with some linear combination method). Alternatively, the may "simply" be an additional source of noise. In the case of AtLAST observations, I have opted to not take the optimal linear combination sensitivity to heuristically account for cosmic backgrounds (e.g. CMB and Cosmic Infrared Background, CIB) and dust.
            %, but detailed investigations of separating backgrounds at $10^{\prime\prime}$ scales is warranted.
            There will also be some floor to the noise achievable: the confusion limit. For the case of SZ observations, previous calculations for confusion limits have focused on what is likely to be found for CMB surveys, with instruments with resolution (FWHM of their beams) worse than 1 arcminute. Recently, \citet{raghunathan2022} found that one may expect a confusion limit $<5\times 10^{-7}$ in Compton $y$ for CMB-HD \citep[FWHM of 0.25 arcminutes at 150 GHz][]{sehgal2019}. In particular, much of the confusion would arise from halos with $M_{200} < 10^{13}$ M$_{\odot}$, which is the expected mass threshold for AtLAST \citep{dimascolo2024}. 

            With these additional noise concerns in mind, it's clear that the constraints on hydrostatic mass bias due to turbulent motions will degrade relative to what was presented in Section~\ref{sec:results} and results presented in Figure~\ref{fig:MZ_sigA2d-thresh} may also degrade. Further study quantifying these additional sources of noise is warranted, but outside the scope of this work.

        %\subsubsection{Tracing of quasi-turbulent motions}

    \subsection{Implications for AtLAST}

        With my assumed mapping speed, AtLAST achieves a sensitivity of $\lesssim 5 \times 10^{-7}$ in Compton $y$ with 1 Ms on source integration time. Assuming a 2 deg$^{2}$ FOV for AtLAST, and a $\sim20$\% telescope downtime over a year, one could survey to this depth over $\sim 250$ deg$^{2}$ over 5 years. \citet{dimascolo2024} finds that this sensitivity could be achieved over 4000 deg$^{2}$ (over 5 years) if the noise adheres to the optimal internal linear combination. Given cluster catalogs from ACT \citep{hilton2018,hilton2021} and SPT \citep{bleem2015,bleem2022}, the number of clusters with $M_{500} \gtrsim 3.5 \times 10^{14} \text{M}_{\odot}$ in $\sim 250$ deg$^{2}$ and 4000 deg$^{2}$ will be $\sim 30$ and $\sim 500$, respectively.

        AtLAST has the potential to excel at high redshift, especially with constraints out to $R_{200}$. AtLAST's beam should not be a hindrance to achieving significant constraints on $\sigma_{b_{\mathcal{M}}}$ at $R_{500}$ and $R_{200}$ for many clusters, but for smaller clusters (i.e. at higher redshifts), it presents a potential limitation.

    \subsection{Impact of results}
    %\textcolor{red}{Maybe some of this to Discussion Section??}

        Non-thermal pressure support may include other (non-turbulent) sources of pressure support which are not negligible. From a cosmological context, the total mass of clusters is of fundamental interest \citep[e.g.][]{allen2011}. Measurements of total mass can be derived from caustics, weak lensing, and CMB lensing. This may be the preferred avenue for cosmological constraints based on cluster abundance.
                
        However, access to the total non-thermal pressure from total mass measurements does not necessarily reveal which mechanisms provide the non-thermal pressure support. There is thus still a interest in assessing the non thermal pressure due to turbulent motions. In addition to the use of SB fluctuations, high resolution X-ray spectroscopy (such as XRISM or NewAthena) can access turbulent velocities and it can do so directly. However, comparable velocity constraints from spectroscopic measurements are more expensive than from SB fluctuations. Therefore, SB fluctuations will remain a critical tool in assessing turbulent gas velocities at large cluster radii, especially at moderate to high redshifts.
        
        %provide empirical data to confirm relations between SB fluctuations and turbulent velocities inferred from those fluctuations. 
        %If the scientific objective is to explicitly constrain (quasi) turbulent motions in the ICM, then the two tools appear to be SB fluctuations and high resolution X-ray spectroscopy. While the latter will directly constrain gas velocities, SB fluctuations is the more economical avenue to constrain turbulent motions, especially at large cluster-centric radii. 

\section{Conclusions}
\label{sec:conclusions}

    In this work I investigated concerns of the accuracy of power spectra estimation on noiseless synthetic data. Separately I utilized properties of variance (and power spectra) to derive estimates of uncertainties in power spectra assuming radial surface brightness profiles and noise profiles. These two investigations laid the groundwork to forecast what constraints can be produced with various current and upcoming SZ and X-ray instruments.

    \textbf{On matters of accuracy.}
        In Section~\ref{sec:accuracy} I revisited known biases and unveiled a secondary manner in which beam convolution (which acts directly on $\delta y$ or $\delta S$ and not $\delta y / \bar{y}$ or $\delta S / \bar{S}$) can bias results. I find that accounting for biases due to the method of power spectrum calculation, i.e. shape and mask biases, especially for the \citep[][;A12]{arevalo2012} method are of primary concern. Conversely, the secondary bias due to beam convolution and projection biases are of less concern and it may be advisable to not attempt to correct for a projection bias.
        
        %The matter of projection biases being of less concern is fortunate as a proper correction may not be viable in the case of a spatially evolving fluctuation spectrum. Moreover, in the case of a spatially evolving fluctuation spectrum, attempting to correct for the projection bias of a single underlying power spectrum may yield an incorrect bias.

    %We have validated that the formalism that allows one to deproject surface brightness fluctuations to 3D thermodynamic fluctuations holds even across large areas of a cluster. That is, even over a range of cluster-centric radii, $\theta$, and consequently over an area where $W(\theta,z)$ changes appreciably, the formalism is valid. Given the idealized simulations performed, the validation is evident when working with FFTs. However, when working with the \citep[][;A12]{arevalo2012} method, one must correct for both spectral shape biases as well as masking biases. In particular, if one does not perform these corrections, the inferred injection scale may be significantly off, while the amplitude at the recovered peak should be close to that of the true peak. Correcting for the biases is itself problematic as obtaining the most accurate correction for these biases effectively requires knowledge of the spectrum that one is attempting to recover. As mentioned in Section~\ref{sec:remarks}, perhaps an iterative approach can solve this, but we leave this for another work which tackles more realistic observational cases.

    \textbf{On matters of uncertainty and choice of regions}
    
    For plausible, circular, surface brightness profiles (and noise or exposure profiles), a minimum in spectral uncertainty can be obtained at $\sim 0.2 R_{500}$.
    Physical expectations and instrumental factors lead me to propose (when possible) a scheme of annuli with boundaries: [0-0.4] $R_{500}$, [0.4-1.0] $R_{500}$, and if data quality allows, [1.0-1.5] $R_{500}$.

    \textbf{Forecasting constraints for various instruments}
    I assume universal surface brightness profiles for both X-ray and SZ observations and non-thermal pressure profiles to derive constraints on amplitude spectra (at various scales) for a handful of current and proposed X-ray and SZ facilities. These uncertainties are propagated to an uncertainty in the hydrostatic mass bias (due to turbulent motions). In this idealized investigation, many instruments are or will be capable of constraining to hydrostatic mass bias to $\sigma_{b_{\mathcal{M}}} = 0.1$, where \textit{Lynx} and AtLAST perform best. 
    %Accounting for requirements to cover a range of scales, the range of clusters where I find that $\sigma_{b_{\mathcal{M}}} = 0.1$ can plausibly achieved is reduced from the highly idealized case.
    Accounting for requiments to sufficiently recover a fluctuation spectrum, I find that the target of $\sigma_{b_{\mathcal{M}}} = 0.1$ on $M_{500}$ can likely be achieved for clusters with $M_{500} \gtrsim 3 \times 10^{14} M_{\odot}$ at $z=0.5$ (and lower masses for $z < 0.5$) by AtLAST and \textit{Lynx}. Both AtLAST and \textit{Lynx} face some challenges to perform constraints down to  $M_{500} \simeq 3 \times 10^{14} M_{\odot}$ at $z>0.5$. AtLAST appears to have a clear advantage in its ability to constrain $\sigma_{b_{\mathcal{M}}}$ on $M_{200}$.
    %However, AtLAST maintains this mass threshold out to larger redshifts, while constraints from \textit{Lynx} degrade towards larger redshifts. Moreover, AtLAST should achieve similar constraints on hydrostatic mass bias at $R_{200}$ for the same mass range.

    %In particular, for 1 Ms of on-source time, AtLAST can achieve $\sigma_{b_{\mathcal{M}}} \simeq 0.1$ on both $b_{\mathcal{M},M_{500}}$ and $b_{\mathcal{M},M_{200}}$ for clusters with  $M_{500} \gtrsim 3 \times 10^{14} \text{M}_{\odot}$. \textit{Lynx} can achieve similar constraints on $b_{\mathcal{M},M_{500}}$ for similar mass clusters at $z=0.5$ ($M_{500} \gtrsim 3 \times 10^{14}$), but can only achieve $\sigma_{b_{\mathcal{M}}} \simeq 0.1$ on $b_{\mathcal{M},M_{200}}$ for clusters with mass $M_{500} \gtrsim 5 \times 10^{14}$ at $z=0.5$. The mass threshold for \textit{Lynx} improves for clusters with $z < 0.5$ and conversely worsens for clusters with $z > 0.5$. My work shows that constraining turbulent pressure support out to $R_{500}$ is expensive for any instrument, and to do so for a sample of clusters, is best studied with an instrument like AtLAST. 

\begin{acknowledgments}

I would like to thank my colleagues for their insights and encouragement in previous works which acted as a springboard for this work. I would like to further thank Max Gaspari and Paul Nulsen for their theoretical insights, Rishi Khatri for sanity checks along the way, Gerrit Schellenberger and Ralph Kraft for their expertise with X-ray observations, and Irina Zhuravleva for the idea to use a map of $\sqrt{N}$. 
CR is supported by NASA ADAP grant 80NSSC19K0574 and Chandra grant G08-19117X.

\end{acknowledgments}

\software{ astropy \citep{astropy2013,astropy2018,astropy2022}, SciPy \citep{SciPy2020}, numpy \citep{Numpy2020}, emcee \citep{foreman2013}, pyproffit \citep{eckert2017}, SOXS \citep{soxs2023} }

\bibliography{references}{}
\bibliographystyle{aasjournal}

\appendix

\section{Formalism with multiple underlying power spectra}
\label{sec:multP3D_formalism}

    \begin{figure}[!h]
        \begin{center}
            \includegraphics[width=0.49\textwidth]{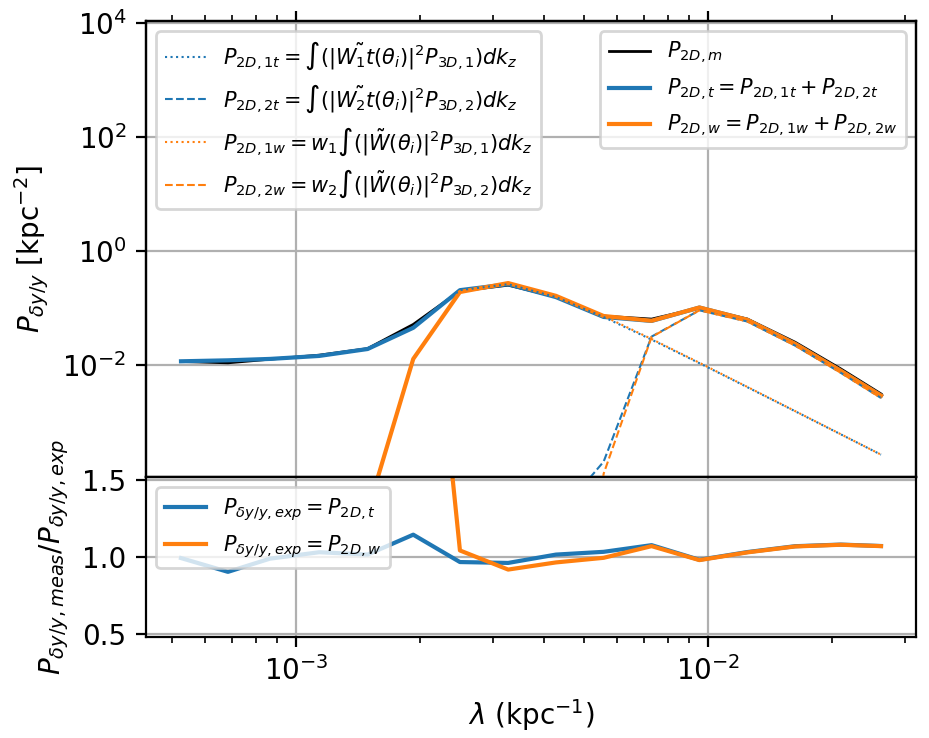}
        \end{center}
        \caption{The resultant $P_{2D}$ spectrum due to two distinct $P_{3D}$ along the line of sight is roughly the weighted sum of the two corresponding $P_{2D}$ (see text). Here, I use $P_{2D,1\text{w}} = w_1 P_{2D,1}$ and $P_{2D,2\text{w}} = w_2 P_{2D,2}$.}
        \label{fig:multipleP3D_numerical_int}
    \end{figure}

    In order to explore the effects of multiple $P_{3D}$ on a resultant $P_{2D}$, I perform a simple numerical simulation wherein I assume two distinct power spectra, $P_{3D,1}$ and $P_{3D,2}$ lie in separate regions along the line of sight. 
    In particular I create a 3D box where a region (e.g. $|z| < 200$ kpc) is occupied by ($\delta P / P$) fluctuations corresponding to $P_{3D,1}$ and the remaining (complementary, e.g. $|z| >= 200$ kpc)) region is occupied by ($\delta P / P$) fluctuations corresponding to $P_{3D,2}$ with the same underlying Gaussian random field. I numerically integrate $W (\delta P / P) $ for $W(\theta,z)$ at a specific $\theta$ ($500$ kpc). To calculate $W$, I assume an A10 profile of a $z=0.3$ and $M_{500} = 6 \times 10^{14} M_{\odot}$ cluster. 

    I adopt a weighting scheme for each $P_{\text{3D}}$, $t(r)$, where $t = 1$ where the fluctuations have the underlying power spectrum $P_{3D,1}$ and $t = 0$ otherwise to transition between the two power spectra, giving:
    \begin{align}
        P_{\text{3D,1t}}(k,r) &= P_{\text{3D,1}}(k)  t(r) \text{, and} \\
        P_{\text{3D,2t}}(k,r) &= P_{\text{3D,2}}(k)  [1 - t(r)]. \\
    \end{align} 
    When considering the projection, we can apply the weights to the window functions $W_{1t}(\theta,z) = W(\theta,z) t(\theta,z)$ and $W_{2t}(\theta,z) = W(\theta,z) [1 - t(\theta,z)]$ such that:
    \begin{align}
        P_{\text{2D},1\text{t}} &= \int P_{\text{3D}} |\tilde{W}_{1\text{t}})^2 dk_z \text{, and} \\
        P_{\text{2D},2\text{t}} &= \int P_{\text{3D}} |\tilde{W}_{2\text{t}})^2 dk_z.
    \end{align}

    It is of no surprise that the discontinuity between the two regions induces aliasing. Thus I alter the weight function from a step function to that of a Tukey window (tapered cosine). Specifically, I found that a width of six resolution elements produced a sufficiently smooth numerical transition; these results can be seen in Figure~\ref{fig:multipleP3D_numerical_int}

    It is also possible to derive (another) approximation in the case of multiple underlying $P_{\text{3D}}$.
    From Equation~\ref{eqn:formalism_full}, I derive $P_{\text{2D},1}$ and $P_{\text{2D},2}$ corresponding to the integration of $P_{\text{3D},1}$ and $P_{\text{3D},2}$, respectively, over $dk_z$ with the same $|\Tilde{W}|^2$. I the following values:
    \begin{align}
        w_1 &= \frac{\int (t W)^2 dz}{\int (W^2) dz} \text{, and} \\
        w_2 &= \frac{\int ([1-t] W)^2 dz}{\int (W^2) dz},
    \end{align}
    respectively, . With these weights, I find:
    \begin{equation}
        P_{2D,\text{w}} \approx w_1 P_{2D,1} + w_2 P_{2D,2}.
    \end{equation}
    Figure~\ref{fig:multipleP3D_numerical_int} shows that the approximate form does quite well, especially for features that are likely of interest.
        %It is not too surprising that this is not fully accurate. In particular, the lack of accuracy appears most prominent at large scales. In the case that $P_{3D,1}$ has larger fluctuations than $P_{3D,2}$ at a given (large) scale, $l$, but $P_{3D,1}$ occupies a region $|z| < x$, where $x \lesssim l$, then it is not surprising that the fluctuations at these scales will fall below their expected amplitude. \textcolor{red}{This approximation should be sufficient for practical use.}

    \subsection{Uncertainty in hydrostatic bias}
    \label{sec:HMBU}

        For each instrument, redshift, cluster mass, injection scale, region, and exposure time, I calculate the inferred uncertainty in the 2D amplitude spectrum as indicated in Section~\ref{sec:uncertainties}. The only bias corrections applied to power spectra are those from the PSF correction, wherein I take $\alpha=3$.\footnote{The choice of $\alpha=3$ is motivated because this should be indicative of the bias around the peak, where (necessarily) $d \ln P_{\text{3D}} / d \ln r = -3$.} I use the regions outlined in Section~\ref{sec:synthesis}: rings with bounds at [0.4, 1.0, 1.5] $R_{500}$; the innermost ring being a circle of radius $0.4 R_{500}$. Values for $N_{\rm eff}$ are calculated as in Section~\ref{sec:accuracy} for each ring and allow me to deproject spectra to $P_{\rm 3D}$ and $\sigma_{P_{\rm 3D}}$.
        
        %Let's write out the uncertainty in $b_{\mathcal{M}}$ as if each variable is independent of the other, e.g. $d \ln \mathcal{M}_{\text{3D}} / d \ln r$ is independent of $\mathcal{M}_{\text{3D}}$ (which it clearly is not).
        To condense expressions, I will adopt the following nomenclatures:
        \begin{align*}
            \alpha_{\text{NT}}   &= d \ln P_{\text{NT}} / d \ln r \\
            \alpha_{\text{th}}   &= d \ln P_{\text{th}} / d \ln r \\
            \alpha_{\mathcal{M}} &= d \ln \mathcal{M}_{\text{3D}} / d \ln r
        \end{align*}
        With the above condensations, I derive:
        \begin{equation}
            \sigma_{\alpha_{\text{NT}}/\alpha_{\text{th}}} = 2 \left[ \left(\frac{\sigma_{\alpha_{\mathcal{M}}}}{\alpha_{\text{th}}} \right)^2 + \left(\frac{\alpha_{\mathcal{M}} \sigma_{\alpha_{\text{th}}}}{\alpha_{\text{th}}^2} \right)^2 \right]^{1/2}
            \label{eqn:NTunc}
        \end{equation}
        While the uncertainties on $\mathcal{M}$ at [0.4, 1.0, 1.5] $R_{500}$ is dependent on $\alpha_{\mathcal{M}}$, I treat them as being independent for simplicity of calculation. The uncertainty of the hydrostatic mass bias can then be computed as:
        \begin{multline}
            \sigma_{b_{\mathcal{M}}} = \frac{\gamma}{3} \left[ \left(2 \mathcal{M}_{\text{3D}} \sigma_{\mathcal{M}} \frac{\alpha_{\text{NT}}}{\alpha_{\text{th}}}\right)^2 + (\mathcal{M}_{\text{3D}}^2 \sigma_{\alpha_{\text{NT}}/\alpha_{\text{th}}})^2 \right]^{1/2} \\ \times \left( 1 +  \frac{\gamma \mathcal{M}_{\text{3D}}^2}{3}\frac{ \alpha_{\text{NT}}}{\alpha_{\text{th}}} \right)^{-2}.
            \label{eqn:bMunc}
        \end{multline}        

        % \frac{ d \ln P_{\text{NT}}}{d \ln P_{\text{th}}}

        %\textcolor{red}{REVISIT THIS: ADDRESS CALCULATION AT $R_{500}$ rather than simply within RING 2.}
        In this computation, the fundamental uncertainties are taken as $\sigma_{\alpha_{\mathcal{M}}}$, $\sigma_{\alpha_{\text{th}}}$, and $\sigma_{\mathcal{M}}$. The uncertainty in the logarithmic slope of the Mach number, $\sigma_{\alpha_{\mathcal{M}}}$ is calculated by fitting a single power law with just two rings (Ring 1 and Ring 2) or all three rings.  The uncertainty in the logarithmic slope of thermal pressure, $\sigma_{\alpha_{\text{th}}}$, is set to zero for X-ray instruments as a robust estimation is beyond the intended scope of this paper. 
        
        For SZ instruments, $\sigma_{\alpha_{\text{th}}}$ is more readily inferred by calculating the uncertainty on $d \ln y / d \ln r = \alpha_y$ using annuli of [0.8,1.0] and [1.0,1.2] times a ring edge, where ring edges are $0.4 R_{500}$, $1.0 R_{500}$, or $1.5 R_{500}$, where I assume $\sigma_{\alpha_{\text{th}}} \approx \sigma_{\alpha_y}$. This approximation is sufficient at moderate radii and increasingly accurate at large radii, as the pressure profile tends towards behaving as a single power law. See Appendix~\ref{sec:SBto3D} for a derivation of this relation between the Compton $y$ profile and electron pressure, $P_e$ profile which is valid when $P_e$ behaves as a single power law at $r > \theta$, for a projected radius $\theta$.

        Lastly, $\sigma_{\mathcal{M}}$ is calculated by propagating errors through the power law fit and corresponding estimation of $\mathcal{M}$ at  [0.4, 1.0, 1.5] $R_{500}$.
        %\begin{equation}
        %    \sigma_{\frac{ d \ln P_{\text{NT}}}{d \ln P_{\text{th}}}} = 2 \left( \frac{ \sigma_{d \ln \mathcal{M}_{\text{3D}} / d \ln r}}{d \ln P_{\text{th}} / d \ln r} + \frac{ d \ln \mathcal{M}_{\text{3D}} / d \ln r \times \sigma_{d \ln P_{\text{th}} / d \ln r}}{(d \ln P_{\text{th}} / d \ln r)^2} \right).
        %\end{equation}

        %\textcolor{red}{Took a break and lost momentum here. Will return to this.}

\section{Power Law deprojection}
%\label{sec:PLint}
\label{sec:SBto3D}

    Other works have shown analytic expressions for interations along the line of sight for emissivity or pressure which is characterized as a power law \citep[e.g.][]{vikhlinin2001a,korngut2011,romero2018}. Here, let us consider the case where the Compton $y$ profile is taken to be a power law, $y = y_0 \theta^{-\alpha_y}$, for projected radii $\theta$. 
    %Here, we examine the case where beyond some (3D) radius, $r$, the electron pressure, $P_e$, is proportional to the radius to some power. 
    From the inverse Abel transform, we have:
    \begin{equation}
        P_e(r) = \frac{m_e c^2}{\sigma_T} \frac{1}{\pi} \int_{\infty}^{r} \frac{dy}{d\theta} \frac{d\theta}{(\theta^2 - r^2)^{1/2}},
    \end{equation}
    where $\theta$ is the projected radius, $m_e$ is the electron mass, $c$ is the speed of light, and $\sigma_T$ is the Thomson cross section. If drop constants, we can write:
    \begin{equation}
        P_e(r) \propto \int_{\infty}^{r} \frac{\theta^{-\alpha_y - 1} d\theta}{(\theta^2 - r^2)^{1/2}}.
    \end{equation}
    This equation can be converted into a form recognized as the Beta function by defining a new variable, $\zeta = r/\theta$. We now derive:
    \begin{equation}
        P_e(r) \propto r^{-\alpha_y - 1} \int_{0}^{1} \frac{\zeta^{\alpha_y} d\zeta}{(1 - \zeta^2)^{1/2}},
    \end{equation}
    where the integral is the Beta function, and is equal to $\Gamma(\alpha_y+1)\Gamma(0.5)/\Gamma(\alpha_y+1.5)$. The salient point is that $P_e(r) \propto r^{-\alpha_y - 1}$, such that the logarithmic slope of the pressure profile is simply the logarithmic slope of the Compton $y$ profile minus 1.

\section{Recovery of power spectra with plausible measurement errors}
\label{sec:ShapeRecovery}

    Section~\ref{sec:approach} outlines that the uncertainties in amplitude spectra, thus in inferred Mach numbers, are calculated assuming that biases are (perfectly) accounted for. In Section~\ref{sec:plausible}, it is shown that this can be done quite well for a spectrum with no measurement uncertainties (only sample variance). 
            
    \begin{figure}[!h]
        \begin{center}
            \includegraphics[width=0.45\textwidth]{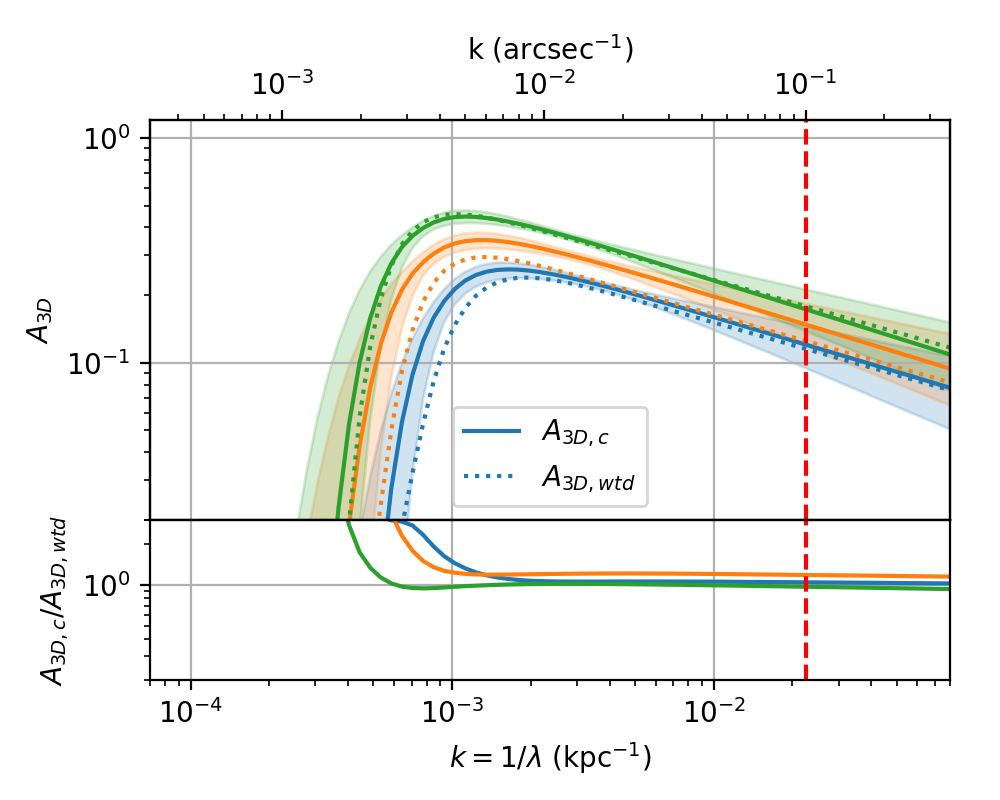}
            \includegraphics[width=0.45\textwidth]{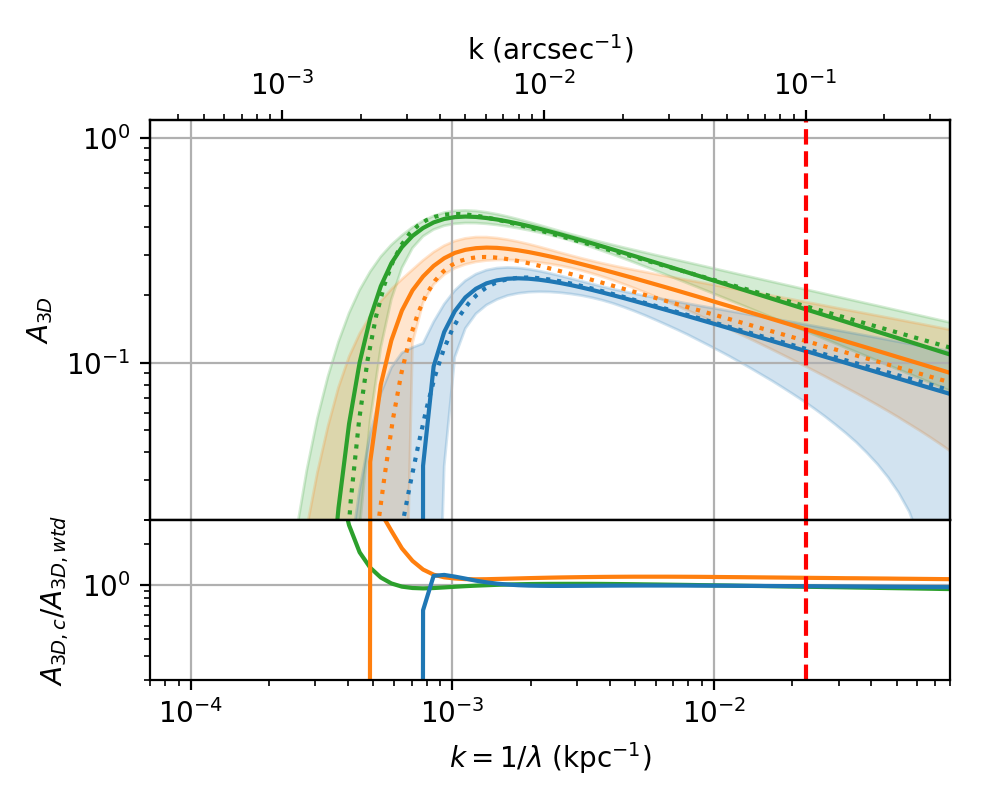}
        \end{center}
        \caption{As in Figure~\ref{fig:RecoveredP3D_withMeasUnc}, but with measurement errors (variance) added to the existing sample variance. In particular, $\sigma_{A_{\text{3D}}}(k_{\text{inj}}*4) = 0.05$, as a plausible target sensitivity in Section~\ref{sec:discussion}.}
        \label{fig:RecoveredP3D_withMeasUnc}
    \end{figure}

    In particular, I explore the option of restricting $\alpha = 11/3$ (or adopting priors such as $\alpha = 11/3 \pm 1/3$), which I consider to be appropriate as this is not only theoretically expected (in most cases), it is well supported observationally \citep[e.g.][]{zhuravleva2018,dupourque2023,heinrich2024}.

    \begin{deluxetable}{ccccc}[!h] 
    \centering 
    \tablecaption{Recovered Amplitude Spectra\label{tbl:as_recovery}} 
    \tablehead{ 
     \colhead{\hspace{0.05cm}Region\hspace{0.05cm}} &\colhead{\hspace{0.5cm} \hspace{0.5cm}} & \colhead{$\ell_{\text{inj}}$ (kpc)} & \colhead{\hspace{0.05cm}$A_{\text{3D}(k_{\text{peak}})}$\hspace{0.05cm}} & \colhead{\hspace{0.05cm}$\sigma_{\text{3D}}$\hspace{0.05cm}}
      } 
    \startdata 
    \multirow{3}{*}{Ring1} & Expected & 515     & 0.24 & 0.33 \\
     & Corrected & 614 & 0.26 & 0.36 \\
     & Peeled & 559 & 0.24 & 0.33 \\
     \hline 
    \multirow{3}{*}{Ring2} & Expected & 746     & 0.30 & 0.40 \\
     & Corrected & 739 & 0.35 & 0.47 \\
     & Peeled & 739 & 0.33 & 0.44 \\
\hline 
    \multirow{2}{*}{Ring3}& Expected & 1003     & 0.46 & 0.58 \\
     & Corrected & 890 & 0.45 & 0.58 \\
\hline 
    \enddata 
    \tablecomments{Properties of recovered A$_{\text{3D}}$ with noise.}
    \vspace{-1cm} 
\end{deluxetable} 

       %it appears to be sufficient to recover the unbiased spectrum quite well where I am chiefly concerned with the accurate recovery of the injection scale and the value of the amplitude spectrum at the injection scale (as demonstrated in Section~\ref{sec:plausible}). 

\section{Expected vs. measured spectral uncertainties}
\label{sec:VerifySpecUnc}

    From Section~\ref{sec:uncertainties}, a relation between a given power spectrum and its uncertainty are given in Equation~\ref{eqn:PS_uncertainty_vs_N}. Additionally, Equation~\ref{eqn:parseval_2D} allows one to convert from the variance (at a specified pixel or beam resolution) to the equivalent white noise power spectrum. 
    
\begin{figure}
        \begin{center}
            \includegraphics[width=0.45\textwidth]{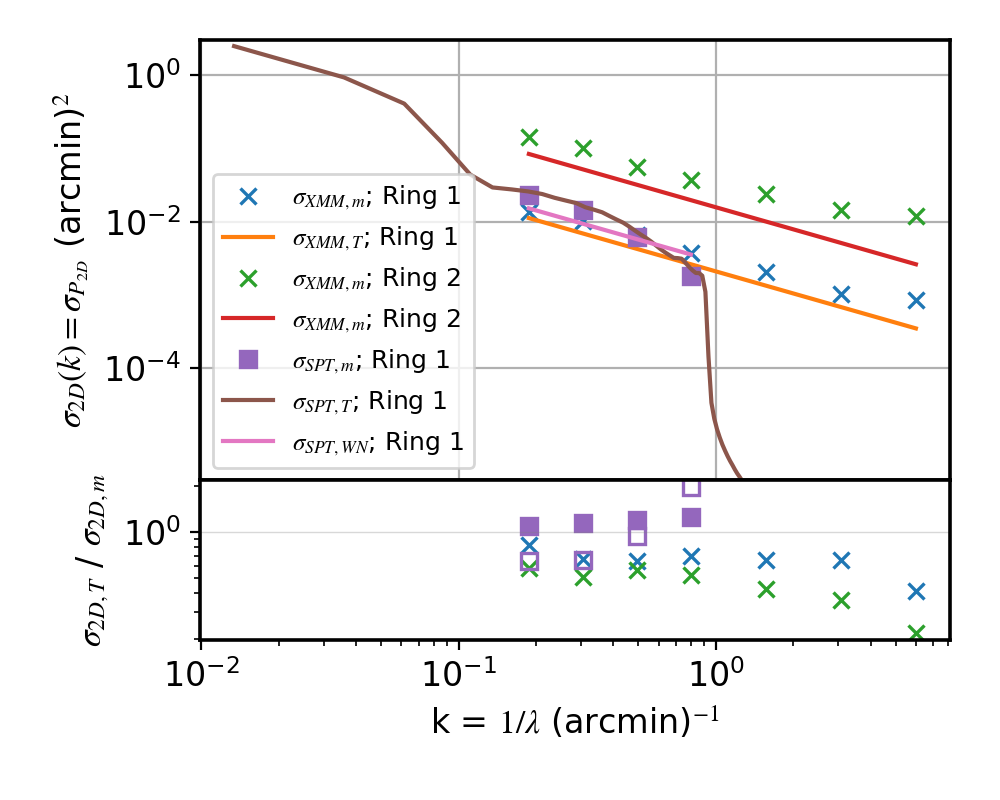}
        \end{center}
        \caption{Verification of Equation~\ref{eqn:PS_uncertainty_vs_N} with data from SPT-CLJ0232-4421 \citep{romero2024}. Both measured and theoretical uncertainties are presented without bias corrections.}
        \label{fig:TheoryVsMeasuredSpectralUncertainty}
    \end{figure}

    Figure~\ref{fig:TheoryVsMeasuredSpectralUncertainty} acts as a check on these relations. The brown curve adopts the measured noise power spectrum \citep{romero2024} and applies Equation\ref{eqn:PS_uncertainty_vs_N}. The remaining "theory" curves (lines) adopt the white noise estimate. Both methods require scaling by the area-average of the square of the inverse (ICM-only) surface brightness model. 

    In the case of deriving the Poisson power spectrum via Equation~\ref{eqn:parseval_2D} for the X-ray case, One finds $k_{\rm max} = k_{\rm pix} / \sqrt{\pi}$, where $k_{\rm pix} = 1 / l_{\rm pix}$ for $l_{\rm pix}$ being the angular length of a pixel (assuming square pixels). This estimation underestimates the reported uncertainties because the reported uncertainties are calculated from realizations which sample the MCMC chain of $\beta$ profile parameters for the given cluster.
    
    In the SZ case, $k_{\rm max} = 1 / (2 \pi \sigma_{\rm PSF}^2) \approx 1/\text{FWHM}^2$ for noise reported at beam scales.

\end{document}